\documentclass[preprint,3p,12pt]{elsarticle}
\usepackage{mathrsfs}
\usepackage{amsmath}
\usepackage{stmaryrd}
\usepackage{bbding}
\usepackage{dcolumn}
\usepackage{graphicx}
\usepackage{amsfonts}
\usepackage{amssymb}
\usepackage{psfrag}
\usepackage{wrapfig}
\usepackage{subfigure}
\usepackage{makeidx}
\usepackage{bm}
\usepackage{epsf}
\usepackage{epsfig}
\usepackage{setspace}
\usepackage{graphicx}
\usepackage{psfrag}
\usepackage{subfigure}
\usepackage{color}
\usepackage{epstopdf}
\begin{document}
\title{Three dimensional high-order gas-kinetic scheme for supersonic isotropic turbulence I: criterion for direct numerical simulation}

\author[HKUST1]{Guiyu Cao}
\ead{gcaoaa@connect.ust.hk}

\author[BNU]{Liang Pan}
\ead{panliang@bnu.edu.cn}

\author[HKUST1,HKUST2,HKUST3]{Kun Xu \corref{cor}}
\ead{makxu@ust.hk}

\address[HKUST1]{Department of Mathematics, Hong Kong University of Science and Technology, Clear Water Bay, Kowloon, Hong Kong}
\address[BNU]{School of Mathematical Sciences, Beijing Normal University, Beijing, China}
\address[HKUST2]{Department of Mechanical and Aerospace Engineering, Hong Kong University of Science and Technology, Clear Water Bay, Kowloon, Hong Kong}
\address[HKUST3]{Shenzhen Research Institute, Hong Kong University of Science and Technology, Shenzhen, China}
\cortext[cor]{Corresponding author}

\begin{abstract}
In this paper, we intend to address the high-order gas-kinetic
scheme (HGKS) in the direct numerical simulation (DNS) of
compressible isotropic turbulence up to the supersonic regime. To
validate the performance of HGKS, the
compressible isotropic turbulence with turbulent Mach number $Ma_t =
0.5$ and Taylor microscale Reynold number $Re_{\lambda}=72$ is
simulated as a benchmark. With the consideration of robustness and accuracy, 
the WENO-Z scheme is adopted for spatial reconstruction in the current higher-order scheme.
Statistical quantities are compared with the high-order compact
finite difference scheme to determine the spatial and temporal criterion for DNS.
According to the grid and time convergence study, it can be
concluded that the minimum spatial resolution parameter
$\kappa_{max} \eta_0 \ge 2.71$ and the maximum temporal resolution
parameter $\Delta t_{ini}/\tau_{t_0} \leq 5.58/1000$ are adequate
for HGKS to resolve the compressible isotropic turbulence, where
$\kappa_{max}$ is the maximum resolved wave number, $\Delta t_{ini}$
is the initial time step, $\eta_0$ and $\tau_{t_0}$ are the initial
Kolmogorov length scale and the large-eddy-turnover time.
Guided by such criterion, the compressible isotropic turbulence from
subsonic regime $Ma_t=0.8$ to supersonic one $Ma_t=1.2$, and the
Taylor microscale Reynolds number $Re_{\lambda}$ ranging from $10$
to $72$ are simulated. With the high initial turbulent Mach
number, the strong random shocklets and high expansion regions
are identified, as well as the wide range of probability density
function over local turbulence Mach number. All those impose great
challenge for high-order schemes. 
In order to construct compressible
large eddy simulation models at high turbulent Mach number,
the ensemble budget of turbulent kinetic energy is fully analyzed. 
The solenoidal dissipation rate decreases with the increasing of
$Ma_t$ and $Re_{\lambda}$. Meanwhile, the dilational dissipation
rate increases with the increasing of $Ma_t$, which cannot be neglected
for constructing supersonic turbulence model. 
The current work shows that HGKS
provides a valid tool for the numerical and physical studies
of isotropic compressible turbulence in supersonic regime, which is
much less reported in the current turbulent flow study. 
\end{abstract}

\maketitle

\section{Introduction}
Compressible turbulence has received great interest for pervading
many important engineering applications and natural phenomena, such
as  hypersonic spacecraft reentry, nuclear fusion power reactors and
interstellar turbulence \cite{aluie2011compressible}. Isotropic
compressible turbulence is regarded as one of cornerstones to
elucidate the effects of compressibility for compressible turbulence
\cite{lele1994compressibility}. Based on the numerical experiments
and theoretical analyses, isotropic compressible turbulence is
divided into four main dynamical regimes \cite{sagaut2008homogeneous}, i.e. the low-Mach number
quasi-isentropic regime, the low-Mach number thermal regime, the
nonlinear subsonic regime, and the supersonic regime. For isotropic incompressible
turbulence in periodic box, the pesudo-spectral method (PSM)
\cite{wang1996examination, moin1998direct} and Lattice-Boltzman
method (LBM) \cite{chen1998lattice, yu2005lattice} have been well
established and applied for incompressible turbulence. However, both
of them are not suitable for compressible turbulence. High-order
compact finite difference method (FDM) \cite{lele1992compact} has
been widely utilized in the simulation of isotropic compressible
turbulence with moderate turbulent Mach number, ranging from the
low-Mach number quasi-isentropic regime to the nonlinear subsonic
regime ($Ma_t \leq 0.8$) \cite{samtaney2001direct,
pirozzoli2004direct, jagannathan2016reynolds}. However, when
simulating the turbulent in supersonic regime ($Ma_t \geq 0.8$), it
fails to capture strong shocklets and suffers from numerical
instability. To study isotropic compressible turbulence with high
turbulent Mach number, the piecewise parabolic method (PPM)
\cite{porter1992numerical, porter1994kolmogorov} has been applied
previously, but the small-scale turbulent structures cannot be
resolved due to excessive numerical dissipation. In this decade,
aiming at capturing shocklets robustly and resolving smooth region
accurately, hybrid scheme combining the compact finite difference
scheme and WENO-type scheme has been developed \cite{adams1996high,
wang2010hybrid}. To the authors' knowledge, due to the instability
when capturing strong shocklets, the biggest turbulent Mach number
of such hybrid scheme has been limited in the critical threshold of
supersonic regime, i.e. $Ma_t \approx 1.0$. For isotropic
compressible turbulence in supersonic regime, the stronger random
shocklets and higher spatial-temporal gradients pose greater
difficulties for numerical analyses than other regimes. Currently,
the supersonic regime is much less known and reported, and only a
very few systematic numerical experiments are available
\cite{wang2012scaling, wang2017scaling, wang2017shocklet,
wang2018effect, wang2018kinetic}.

In the past decades, the gas-kinetic scheme (GKS) based on the
Bhatnagar-Gross-Krook (BGK) model
\cite{bhatnagar1954model,chapman1990mathematical} has been developed
systematically for the computations from low speed flow to
supersonic one \cite{xu1998gas,xu2001gas,xu2015direct}. Different
from the numerical methods based on the macroscopic governing
equations, the gas-kinetic scheme presents a gas evolution process
from kinetic scale to hydrodynamic scale, where both inviscid and
viscous fluxes are recovered from the moments of a single
time-dependent gas distribution function \cite{xu2001gas,
xu2015direct}. In discontinuous shock region, the kinetic scale
physics takes effect to construct a crisp and stable shock
transition. In smooth flow region, the hydrodynamic scale physics
corresponding to the multi-dimensional central difference
discretization will contribute mainly in the kinetic flux function,
and accurate Navier-Stokes solution can be obtained once the flow
structure is well resolved. Both normal and tangential gradients of
flow variables are included in the flux function across a cell
interface \cite{li2010high,luo2013high}. With the discretization of
particle velocity space, a unified gas-kinetic scheme (UGKS) has
been developed for the flow in the entire Knudsen number regimes
from rarefied to continuum ones
\cite{xu2010unified,huang2012unified}. With the two-stage temporal
discretization for the Lax-Wendroff type flow solvers
\cite{li2016two, pan2016efficient}, a reliable framework was
provided for developing the GKS into fourth-order and even
higher-order accuracy with the implementation of the traditional
second-order or third-order flux functions
\cite{pan2017few,ji2018family}. More importantly, this scheme is as
robust as the second-order scheme and works perfectly from the
subsonic to the hypersonic viscous heat conducting flows
\cite{cao2018hmgks}. In comparison with Riemann solver based CFD
methods, the robustness is solely due to the dynamical evolution
model of the time dependent flux function. For the higher-order
schemes, it seems that a reliable physical evolution model becomes
more important due to the absence of large numerical dissipation in
the second-order schemes, and the delicate flow structures captured
in higher-order schemes depend on the quality of the solvers greatly
\cite{ji2018family}. In recent years, GKS has been applied in
turbulence simulation successfully.  For high-Reynolds number
turbulent flows, the second-order and third-order GKS coupled with
traditional eddy-viscosity turbulence models
\cite{jiang2012implicit, righi2016gas, tan2018gas, cao2018igkst}
have been developed and implemented in turbulent flow simulations,
where a newly turbulent collision time $\tau_t$ was defined to model
the turbulent behavior on unresolved grids. As for low-Reynolds
number turbulent flows, the GKS based on modified WENO
reconstruction \cite{liao2009gas, kumar2013weno} have been
implemented in direct numerical simulation (DNS) of decaying isotropic compressible turbulence.
Numerical results confirm the great advantage of GKS in high-speed
flow simulation. Recently, with the two-stage temporal
discretization and WENO reconstruction
\cite{liu1994weighted,jiang1996efficient,borges2008improved}, the
high-order GKS (HGKS) is constructed for simulating
three-dimensional flows \cite{pan2018two}. Numerical results show
the capability to simulate the complicated flows, such as the
isotropic compressible turbulence.

In this paper, we concentrate on the DNS of compressible isotropic
turbulence with high turbulent Mach number, and the two-stage
fourth-order gas-kinetic scheme \cite{pan2016efficient} is adopted
to simulate the compressible isotropic turbulence up to supersonic
regime. In the previous studies, the high resolution can be obtained
only in space. However, high-order accuracy in time is also
necessarily required for DNS to fully resolve the smallest eddies in
turbulent flows, i.e. the eddies in Kolmogorov length scale and time
scale. As a first attempt, the validation of fourth-order GKS for
compressible isotropic turbulence is undertaken to obtain the
criterion to guide the subsequent studies. The current study
indicates that the minimum spatial resolution parameter
$\kappa_{max} \eta_0 \ge 2.71$ and the maximum temporal resolution
parameter $\Delta t_{ini}/\tau_{t_0} \leq 5.58/1000$ are adequate
for the HGKS to resolve the isotropic compressible turbulence. With
the increasing of initial turbulent Mach number, the range of 
the probability density function (PDF) over local turbulence Mach number
becomes wide at the same fixed normalized time. In addition,
stronger random shocklets and higher expansion regions are observed
with the higher initial turbulent Mach number, which exert great
difficulties for high-order schemes. Statistical quantities are
presented for these cases, which are used as the benchmark for
supersonic isotropic turbulence. The solenoidal dissipation rate is
higher with the higher $Ma_t$ and $Re_{\lambda}$. At the same time,
it is observed that the dilational dissipation rate increases 
with the increasing of $Ma_t$, and seems slightly dependent on
$Re_{\lambda}$. This analysis lays foundation for 
constructing compressible large eddy simulation (LES) in supersonic
regime.  This study confirms that HGKS provides a valid tool for
the studies of complex compressible turbulent flows.

This paper is organized as follows. In Section 2, a brief review on
the fourth-order GKS will be presented. Section 3 presents the
detailed flow conditions and statistical turbulence quantities for
isotropic compressible turbulence. Numerical validation and
discussions will be presented in Section 4. Conclusions are shown in
the final section.

\section{Two-stage fourth-order gas-kinetic scheme}
The three-dimensional BGK equation \cite{bhatnagar1954model,chapman1990mathematical} can be
written as
\begin{equation}\label{bgk}
f_t+uf_x+vf_y+wf_z=\frac{g-f}{\tau},
\end{equation}
where $(u,v,w)$ is the particle velocity, $f$ is the gas
distribution function, $g$ is the three-dimensional Maxwellian
distribution, and $\tau$ is the collision time. The collision term
satisfies the compatibility condition
\begin{equation}\label{compatibility}
\int \frac{g-f}{\tau}\psi \text{d}\Xi=0,
\end{equation}
where $\psi=(\psi_1,...,\psi_5)^T=(1,u,v,w,\displaystyle
\frac{1}{2}(u^2+v^2+w^2+\xi^2))^T$, the internal variables $\xi^2$
equals to $\xi^2=\xi_1^2+...+\xi_K^2$,
$\text{d}\Xi=\text{d}u\text{d}vd\text{d}w\text{d}\xi^1...\text{d}\xi^{K}$,
$K$ is the degrees of freedom, and the specific heat ratio
$\gamma=(K+5)/(K+3)$ for three-dimensional flows. Based on the
Chapman-Enskog expansion, the Euler and Navier-Stokes equations can
be derived \cite{xu2015direct,xu2001gas}.

Taking conservative moments of Eq.\eqref{bgk} and integrating over the control
volume $V_{ijk}=\overline{x}_i\times\overline{y}_j\times
\overline{z}_k$ with $\overline{x}_i=[x_i-\Delta x/2,x_i+\Delta
x/2], \overline{y}_j=[y_j-\Delta y/2,y_j+\Delta y/2],
\overline{z}_k=[z_k-\Delta z/2,z_k+\Delta z/2]$,  the
semi-discretized finite volume scheme can be written as
\begin{align}\label{finite}
\frac{\text{d}Q_{ijk}}{\text{d}t}=\mathcal{L}(Q_{ijk})=\frac{1}{\Delta
x\Delta y\Delta z}\Big[
&\int_{\overline{y}_j\times\overline{z}_k}(F_{i-1/2,j,k}-F_{i+1/2,j,k})\text{d}y\text{d}z\nonumber\\
+&\int_{\overline{x}_i\times\overline{z}_k}(G_{i,j-1/2,k}-G_{i,j+1/2,k})\text{d}x\text{d}z\nonumber\\
+&\int_{\overline{x}_i\times\overline{y}_j}(H_{i,j,k-1/2}-H_{i,j,k+1/2})\text{d}x
\text{d}y\Big],
\end{align}
where $Q=(\rho,\rho U,\rho V,\rho W,\rho E)^T$ are the conservative
flow variables, $Q_{ijk}$ is the cell averaged value over the
control volume $V_{ijk}$. For the direct numerical simulation of the
compressible isotropic turbulence, the semi-discretized finite
volume scheme Eq.\eqref{finite} needs to be fully discretized with
high-order accuracy. Recently, a two-stage fourth-order
time-accurate discretization has been  developed for Lax-Wendroff type
flow solvers \cite{li2016two,pan2016efficient}, which provides a reliable
framework to develop high-order scheme for three-dimensional flows
with complicated flow structure. Consider the following
time-dependent equation
\begin{align*}
\frac{\text{d} Q}{\text{d}t}=\mathcal {L}(Q),
\end{align*}
with initial condition
\begin{align*}
Q(t=t_n)=Q^n,
\end{align*}
where $\mathcal {L}$ is an operator for spatial derivative of flux
given by Eq.\eqref{finite}, and the subscript of $Q_{ijk}$ is
omitted for simplicity. The state $Q^{n+1}$ is updated with the
following formula
\begin{equation}\label{two-stage}
\begin{aligned}
&Q^*=Q^n+\frac{1}{2}\Delta t\mathcal {L}(Q^n)+\frac{1}{8}\Delta
t^2\frac{\partial}{\partial
t}\mathcal{L}(Q^n),\\
Q^{n+1}=&Q^n+\Delta t\mathcal {L}(Q^n)+\frac{1}{6}\Delta
t^2\big(\frac{\partial}{\partial
t}\mathcal{L}(Q^n)+2\frac{\partial}{\partial
t}\mathcal{L}(Q^*)\big).
\end{aligned}
\end{equation}
It can be proved that for hyperbolic equations the above time
stepping method Eq.\eqref{two-stage} provides a fourth-order time
accurate solution for $Q(t)$ at $t=t_n +\Delta t$
\cite{li2016two,pan2016efficient}.

To achieve the high-order spatial accuracy, the Gaussian quadrature
for the numerical flux is used at the cell interface. For example,
the numerical flux in $x$-direction is given as
\begin{align}\label{gauss}
\int_{\overline{y}_j\times\overline{z}_k}F_{i+1/2,j,k}\text{d}y\text{d}z=\Delta
y\Delta z\sum_{m,n=1}^2\omega_{mn}
F(\boldsymbol{x}_{i+1/2,j_m,k_n},t),
\end{align}
where $\omega_{mn}$ is the quadrature weight,
$\boldsymbol{x}_{i+1/2,m,n}=(x_{i+1/2},y_{j_m},z_{k_n})$,
$(y_{j_m},z_{k_n})$ is the Gauss quadrature point of the cell
interface $\overline{y}_j\times\overline{z}_k$, and the numerical
flux $F(\boldsymbol{x}_{i+1/2,j_m,k_n},t)$ is provided by taking
moments of the gas distribution function
\begin{align}\label{flux}
F(\boldsymbol{x}_{i+1/2,j_m,k_n},t)=\int\psi u
f(\boldsymbol{x}_{i+1/2,j_m,k_n},t,\boldsymbol{u},\xi)\text{d}u\text{d}v\text{d}w\text{d}\xi.
\end{align}
For the three-dimensional flows, the gas distribution function at
the Gauss quadrature point is given by the second-order gas-kinetic
solver as follows
\begin{align*}
f(\boldsymbol{x}_{i+1/2,j_m,k_n},t,\boldsymbol{u},\xi)=&(1-e^{-t/\tau})g_0+((t+\tau)e^{-t/\tau}-\tau)(\overline{a}_1u+\overline{a}_2v+\overline{a}_3w)g_0\\
+&(t-\tau+\tau e^{-t/\tau}){\bar{A}} g_0\\
+&e^{-t/\tau}g_r[1-(\tau+t)(a_{1r}u+a_{2r}v+a_{3r}w)-\tau A_r)]H(u)\\
+&e^{-t/\tau}g_l[1-(\tau+t)(a_{1l}u+a_{2l}v+a_{3l}w)-\tau
A_l)](1-H(u)).
\end{align*}
In order to implement the two-stage method, Eq.\eqref{flux} is
approximated by a linear function
\begin{align*}
F(\boldsymbol{x}_{i+1/2,j_m,k_n},t)\approx\underbrace{F_{i+1/2,j,k}(Q^n,t_n)}_{\mathcal
{L}}+\underbrace{\partial_t
F_{i+1/2,j,k}(Q^n,t_n)}_{\mathcal{L}_t}t.
\end{align*}
More details for the implementation of gas-kinetic scheme can be
found in \cite{pan2016efficient,pan2017few}.

To achieve the high-order spatial accuracy, the fifth-order WENO
reconstruction \cite{liu1994weighted, jiang1996efficient, borges2008improved} is adopted at the
Gaussian quadrature points. The one-dimensional WENO scheme is given
as follows
\begin{align*}
Q_i^{r}=\sum_{k=0}^2\omega_kQ_i^{kr},~~Q_i^{l}=\sum_{k=0}^2\widetilde{\omega}_kQ_i^{kl},
\end{align*}
where $Q_i^{kr}$ and $Q_i^{kl}$ are obtained by the third-order
interpolation, and $\omega_{k}$ is the nonlinear weight. The
nonlinear weights of WENO-JS \cite{jiang1996efficient} and WENO-Z \cite{borges2008improved}
scheme are given as follows
\begin{align*}
&\omega^{JS}_{k}=\frac{\alpha^{JS}_{k}}{\sum_{m=0}^{2}\alpha^{JS}_{m}},~~\alpha^{JS}_{k}=\frac{d_k}{\beta_k+\varepsilon},\\
\omega^{Z}_{k}&=\frac{\alpha^{Z}_{k}}{\sum_{m=0}^{2}\alpha^{Z}_{m}},~~\alpha^{Z}_{k}=d_k\Big[1+\big(\frac{\tau}{\beta_k+\varepsilon}\big)\Big],
\end{align*}
where  $d_k$ is the linear weights and $\beta_k$ is the smooth
indicator for each candidate stencil.

In the high-order gas-kinetic scheme, the conservative variables
$Q_{0}$ can be determined according to the compatibility condition
Eq.\eqref{compatibility}
\begin{align*}
\int\psi g_{0}\text{d}\Xi=Q_0=\int_{u>0}\psi
g_{l}\text{d}\Xi+\int_{u<0}\psi g_{r}\text{d}\Xi.
\end{align*}
where $g_l$ and $g_r$ are the equilibrium states corresponding to
the conservative variables $Q^r_{i}$ and $Q_{i+1}^l$ at the cell
interface. With the reconstructed variables, the normal spatial
derivatives for the conservative variables at left side, right side
and across the cell interface can be given as follows
\begin{align*}
\partial_xQ_{l}=(Q^r_{i}-Q^l_{i})/\Delta x,~
\partial_xQ_{r}=(Q^r_{i+1}-Q^l_{i+1})/&\Delta x, \\
\partial_xQ_{0}=\big[-\frac{1}{12}(Q_{i+2}-Q_{i-1})+\frac{5}{4}(Q_{i+1}-Q_{i})\big]/&\Delta
x.
\end{align*}
With the reconstructed conservative variables and normal derivatives
in normal direction, the point value $Q_l, Q_r$ and $Q_0$ and
first-order derivatives at the Gauss quadrature points
$\boldsymbol{x}_{i+1/2,m,n}=(x_{i+1/2},y_{j_m},z_{k_n})$ can be
constructed. The detailed procedure is given as follows
\begin{enumerate}
\item According to one dimensional reconstruction, the cell averaged reconstructed values and cell averaged spatial derivatives
\begin{align*}
&(Q_{l})_{j-\ell_1,k-\ell_2},
(Q_{r})_{j-\ell_1,k-\ell_2},(Q_{0})_{j-\ell_1,k-\ell_2},\\
(\partial_x&Q_{l})_{j-\ell_1,k-\ell_2},
(\partial_xQ_{r})_{j-\ell_1,k-\ell_2},(\partial_xQ_{0})_{j-\ell_1,k-\ell_2},
\end{align*}
can be constructed, where $\ell_1,\ell_2=-2,...,2$.
\item With the one-dimensional WENO reconstruction in the horizontal direction, the averaged
value and the averaged spatial derivatives
\begin{align*}
(Q_{l})_{j_m,k-\ell_2},
(&Q_{r})_{j_m,k-\ell_2},(Q_{0})_{j_m,k-\ell_2},\\
(\partial_xQ_{l})_{j_m,k-\ell_2}, (&\partial_xQ_{r})_{j_m,k-\ell_2},(\partial_xQ_{0})_{j_m,k-\ell_2},\\
(\partial_yQ_{l})_{j_m,k-\ell_2},
(&\partial_yQ_{r})_{j_m,k-\ell_2},(\partial_yQ_{0})_{j_m,k-\ell_2}
\end{align*}
over the interval $[z_{k-\ell_2}-\Delta z/2,z_{k-\ell_2}+\Delta
z/2]$ with $y=y_{j_m}$ can be given.
\item With one-dimensional WENO reconstruction in the vertical direction, the point value and spatial derivatives
\begin{align*}
(Q_{l})_{j_m,k_n}, &(Q_{r})_{j_m,k_n},(Q_{0})_{j_m,k_n},\\
(\partial_xQ_{l})_{j_m,k_n}, &(\partial_xQ_{r})_{j_m,k_n},(\partial_xQ_{0})_{j_m,k_n},\\
(\partial_yQ_{l})_{j_m,k_n}, &(\partial_yQ_{r})_{j_m,k_n},(\partial_yQ_{0})_{j_m,k_n},\\
(\partial_zQ_{l})_{j_m,k_n},
&(\partial_zQ_{r})_{j_m,k_n},(\partial_zQ_{0})_{j_m,k_n},
\end{align*}
can be fully determined at the Gaussian quadrature points
$\boldsymbol{x}_{i+1/2,m,n}=(x_{i+1/2},y_{j_m},z_{k_n})$.

\textbf{Remark:} For the tangential reconstruction of $Q_{0}$, the
fourth-order polynomials are constructed at the horizontal and
vertical direction. The variables and spatial derivatives can be
constructed at the Guassian quadrature points.

For the tangential reconstruction of $Q_{l,r}$, the variables at
the ends of cell interface can be obtained from the fifth-order WENO
method at the horizontal and vertical direction. With reconstructed
variables and the cell averaged variables, the quadratic polynomials
can be constructed. The variables and spatial derivatives can be
constructed at the Guassian quadrature points as well.

In the part of code validation, the smooth flow fields without
strong shocklets at $Ma_{t} = 0.1, 0.3$ and $0.5$ are calculated first.
The simplified smooth second-order gas-kinetic flux
\cite{pan2018two} and WENO scheme with linear weights (WENO-L) are 
used in the validation. To improve the robustness without losing too much accuracy,
for the isotropic turbulence from subsonic to supersonic regime,
i.e. $Ma\geq0.5$, the variable $Q_{0}$ takes the identical tangential
reconstruction of $Q_{l,r}$ in WENO-JS \cite{jiang1996efficient}
and WENO-Z \cite{borges2008improved} scheme.
\end{enumerate}
More details of three-dimensional high-order gas-kinetic scheme can
be found in \cite{pan2018two}.

\section{Decaying isotropic compressible turbulence}
The isotropic compressible turbulence is regarded as one of
fundamental benchmarks to study the compressible effect. Both forced
isotropic compressible turbulence with solenoidal and dilational
external force \cite{jagannathan2016reynolds, kida1990energy,
kida1990enstrophy} and decaying isotropic compressible turbulence
\cite{moin1998direct, samtaney2001direct, pirozzoli2004direct} are
studied in the literature. In this paper, we concentrate on the decaying
isotropic compressible turbulence without external force. The flow
domain of numerical simulation is a cube box defined as $[-\pi, \pi]
\times [-\pi, \pi] \times [-\pi, \pi]$, with periodic boundary
conditions in all three Cartesian directions for all the flow
variables. Evolution of this artificial system is determined by
initial thermodynamic quantities and two dimensionless parameters,
i.e. the initial Taylor microscale Reynolds number
\begin{align*}
Re_{\lambda}=\frac{\left\langle \rho \right\rangle u_{rms}
\lambda}{\left\langle \mu \right\rangle},
\end{align*}
and turbulent Mach number
\begin{align*}
Ma_t=\frac{\sqrt{3} u_{rms}}{\left\langle  c_s \right\rangle},
\end{align*}
where $\left\langle \cdot \right\rangle$ is the ensemble over the
whole computational domain, $\rho$ is the density, $\mu$ is the
initial dynamic viscosity, $c_s$ is the sound speed and $u_{rms}$ is
the root mean square of initial turbulent velocity field
\begin{align*}
u_{rms}=\left\langle \frac{\bm{u} \cdot \bm{u}}{3}
\right\rangle^{1/2}.
\end{align*}
A three-dimensional solenoidal random initial velocity field
$\bm{u}$ can be generated by a specified spectrum, which is given by
\cite{passot1987numerical}
\begin{align}\label{initial_spectrum}
E(\kappa) = A_0 \kappa^4 \exp (-2\kappa^2/\kappa_0^2),
\end{align}
where $A_0$ is a constant to get a specified initial kinetic energy,
$\kappa$ is the wave number, $\kappa_0$ is the wave number at which
the spectrum peaks. In this paper, fixed $A_0$ and $\kappa_0$  in
Eq.(\ref{initial_spectrum}) are chosen for all cases, which are
initialized by $A_0 = 0.00013$ and $ \kappa_0 = 8$. Initial
strategies play an important role in isotropic compressible
turbulence simulation \cite{samtaney2001direct}, especially for the
starting fast transient period during which the divergence of the
velocity increases rapidly and the negative temperature or pressure
often appear. In the computation, the initial pressure $p_0$, density $\rho_0$ and
temperature $T_0$ are set as constant. In this way, the initial Taylor
microscale Reynolds number $Re_{\lambda}$ and turbulent Mach number
$Ma_{t}$ can be determined by
\begin{align*}
Re_{\lambda}=&\frac{(2 \pi)^{1/4}}{4} \frac{\rho_0}{\mu_0}\sqrt{2
A_0}\kappa_0^{3/2},\\
Ma_{t}&=\frac{\sqrt{3}}{\sqrt{\gamma R T_0}}u_{rms},
\end{align*}
where the initial density $\rho_0 = 1$, $\mu_0, T_0$ can be
determined by $Re_{\lambda}$ and $Ma_{t}$ and $\gamma = 1.4$ is the
specific heat ratio. In the simulation, the dynamic velocity is
given by
\begin{align}\label{muuu}
\mu=\mu_0(\frac{T}{T_0})^{0.76}.
\end{align}
With current initial strategy, the initial ensemble turbulent
kinetic energy $K_0$, ensemble enstrophy $\Omega_0$, ensemble
dissipation rate $\varepsilon_0$, large-eddy-turnover time
$\tau_{t_0}$, Kolmogorov length scale $\eta_0$, and the Kolmogorov
time scale $\tau_0$ are given as
\begin{equation}\label{initial_def}
\begin{aligned}
K_0=&\frac{3A_0}{64} \sqrt{2 \pi} \kappa_0^5, ~ \Omega_0=\frac{15
A_0}{256} \sqrt{2 \pi} \kappa_0^7,~
\tau_{t_0}=\sqrt{\frac{32}{A_0}}(2 \pi)^{1/4} \kappa_0^{-7/2},\\
&\varepsilon_0=2\frac{\mu_0}{\rho_0} \Omega_0, ~
\eta_0=(\nu_0^3/\varepsilon_0)^{1/4}, ~
\tau_0=(\nu_0/\varepsilon_0)^{1/2}.
\end{aligned}
\end{equation}
For decaying compressible isotropic turbulence, the local turbulent Mach
number $M_{loc}$, root-mean-square density fluctuations $\rho_{rms}$,
and turbulent kinetic energy $K$ are defined as
\begin{equation}\label{rho_k}
\begin{aligned}
M_{loc}&=\frac{ \bm{u} \cdot \bm{u}}{c_{loc}}, \\
\rho_{rms}&=\sqrt{\left\langle \rho -  \left\langle  \rho \right\rangle \right\rangle}, \\
K&=\frac{1}{2}\left\langle  \rho \bm{u} \cdot \bm{u} \right\rangle,
\end{aligned}
\end{equation}
where $c_{loc}$ is the local sound speed. Starting from the initial flows,
the large eddies transfer their turbulent kinetic energy successively to
smaller  eddies. In this process, the evolution of
turbulent kinetic energy is of interest since it is a fundamental
benchmark for incompressible and compressible turbulence modeling
\cite{yoshizawa1985statistically, pope2001turbulent,
chai2012dynamic}. In this study, the ensemble budget of turbulent
kinetic energy is computed and analyzed briefly, as the decay of the
ensemble turbulent kinetic energy can be described approximately by
\cite{sarkar1991analysis}
\begin{equation}\label{dkdt}
\begin{aligned}
\frac{\text{d}\left\langle K\right\rangle}{\text{d}t}=\varepsilon&+\left\langle p \theta \right\rangle,\\
\varepsilon=\varepsilon_s+&\varepsilon_d,
\end{aligned}
\end{equation}
where $\varepsilon_s=\left\langle \mu \omega_i
\omega_i\right\rangle$ is the ensemble solenoidal dissipation rate,
$\displaystyle\varepsilon_d= \left\langle\frac{4}{3}\mu
\theta^2\right\rangle$ is the ensemble dilational dissipation rate,
$\left\langle p \theta \right\rangle$ is the ensemble
pressure-dilation transfer, $\displaystyle\omega_i=\epsilon_{ijk}
\frac{\partial u_k}{\partial x_j}$ is the fluctuating vorticity,
$\epsilon_{ijk}$ is the alternating tensor and $\theta = \nabla
\cdot \bm{u}$ is the fluctuating divergence of velocity.

\section{Numerical simulation and discussions}
In this section, numerical  simulation and discussions for 
isotropic compressible turbulence will be presented. In all
simulations, the collision time $\tau$ takes
\begin{align*}
\tau = \frac{\mu}{p} + C \frac{|p_L - p_R|}{|p_L + p_R|} \Delta t,
\end{align*}
where $\mu$ is the viscous coefficient obtained from
Eq.\eqref{muuu}, $p_L$ and $p_R$ denote the pressures on the left
and right hand sides at the cell interface. The collision time reduces to $\tau
=\mu /p$ in the smooth flow region. The constant $C$ takes $1.5$ in
the computation, and $\Delta t$ is the time step determined
according to the CFL condition.

\begin{figure}[!h]
\centering
\includegraphics[width=0.49\textwidth]{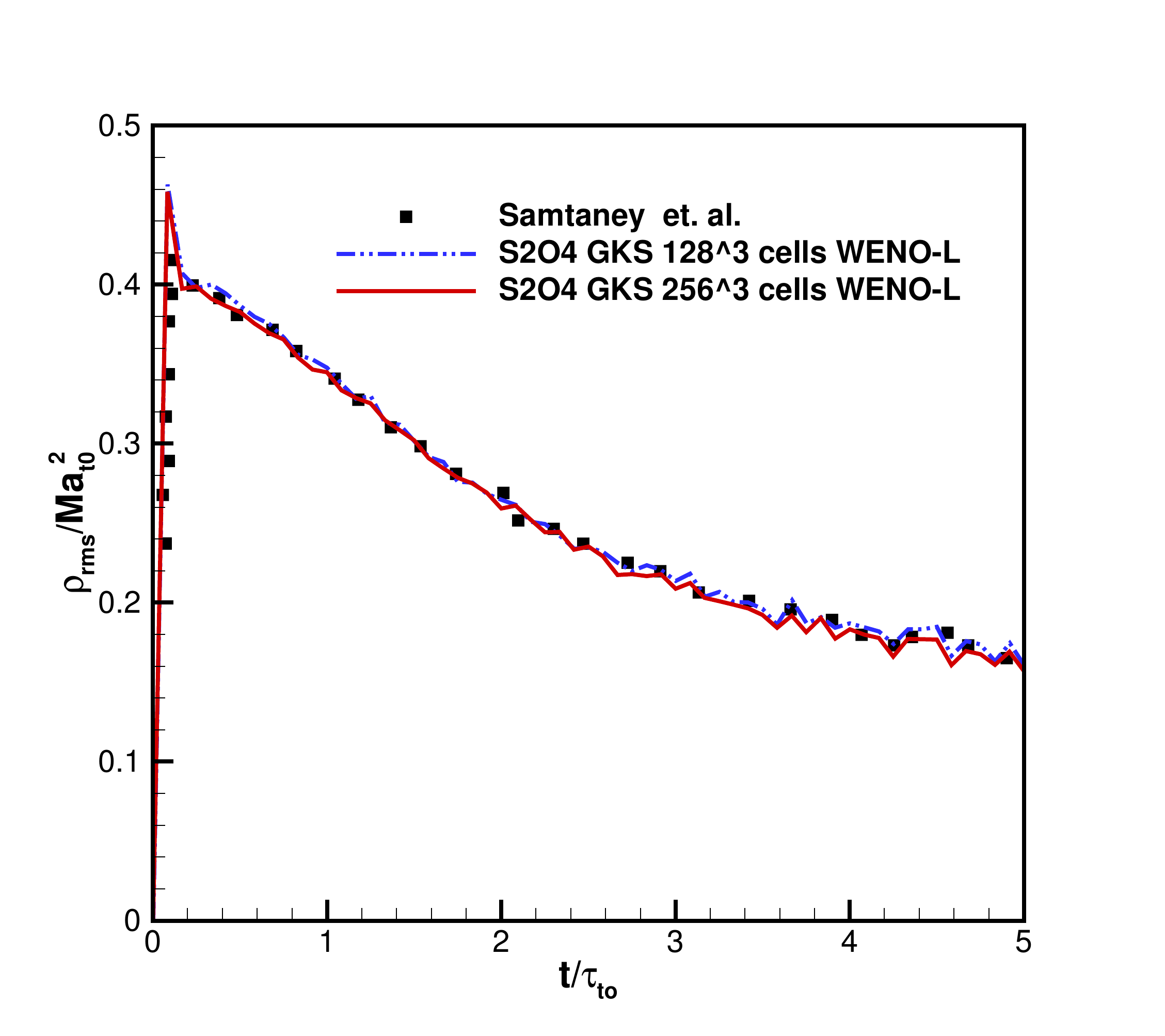}
\includegraphics[width=0.49\textwidth]{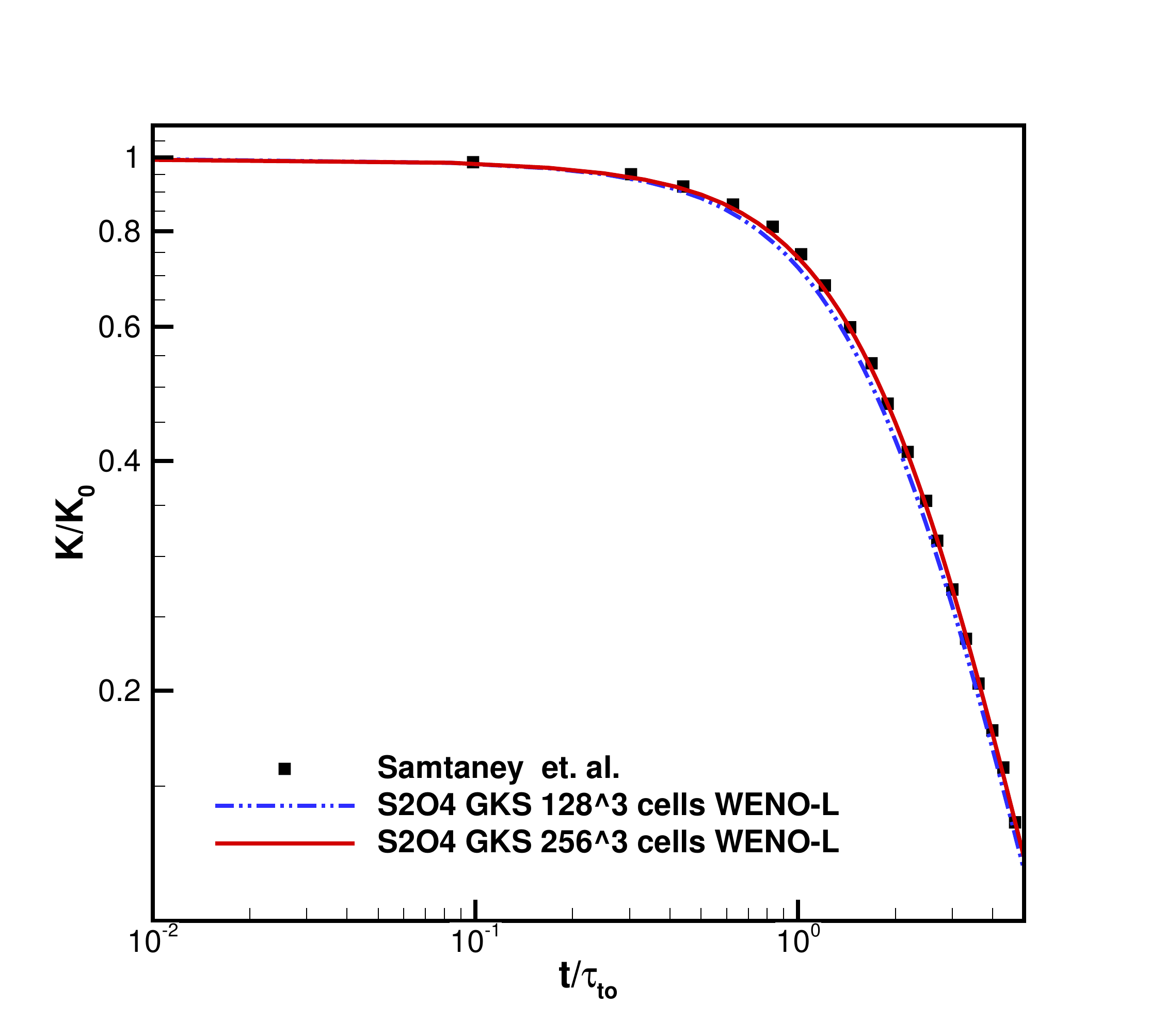}
\caption{\label{Ma01} Time history of $\rho_{rms}/Ma_t^2$ and $K/K_0$ for the near incompressible isotropic turbulence with $Re_{\lambda}=72$ and $Ma_{t}=0.1$.}
\includegraphics[width=0.49\textwidth]{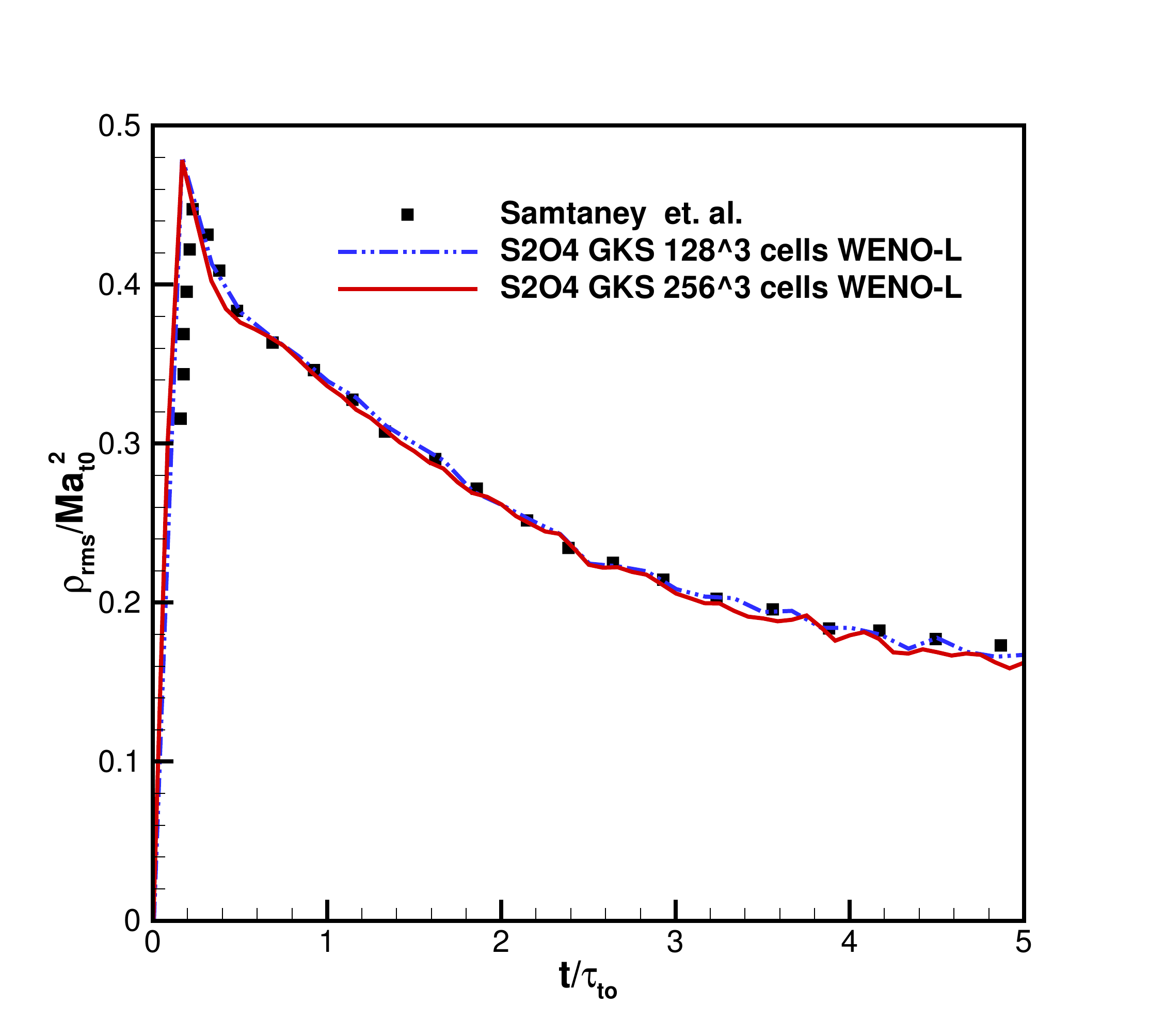}
\includegraphics[width=0.49\textwidth]{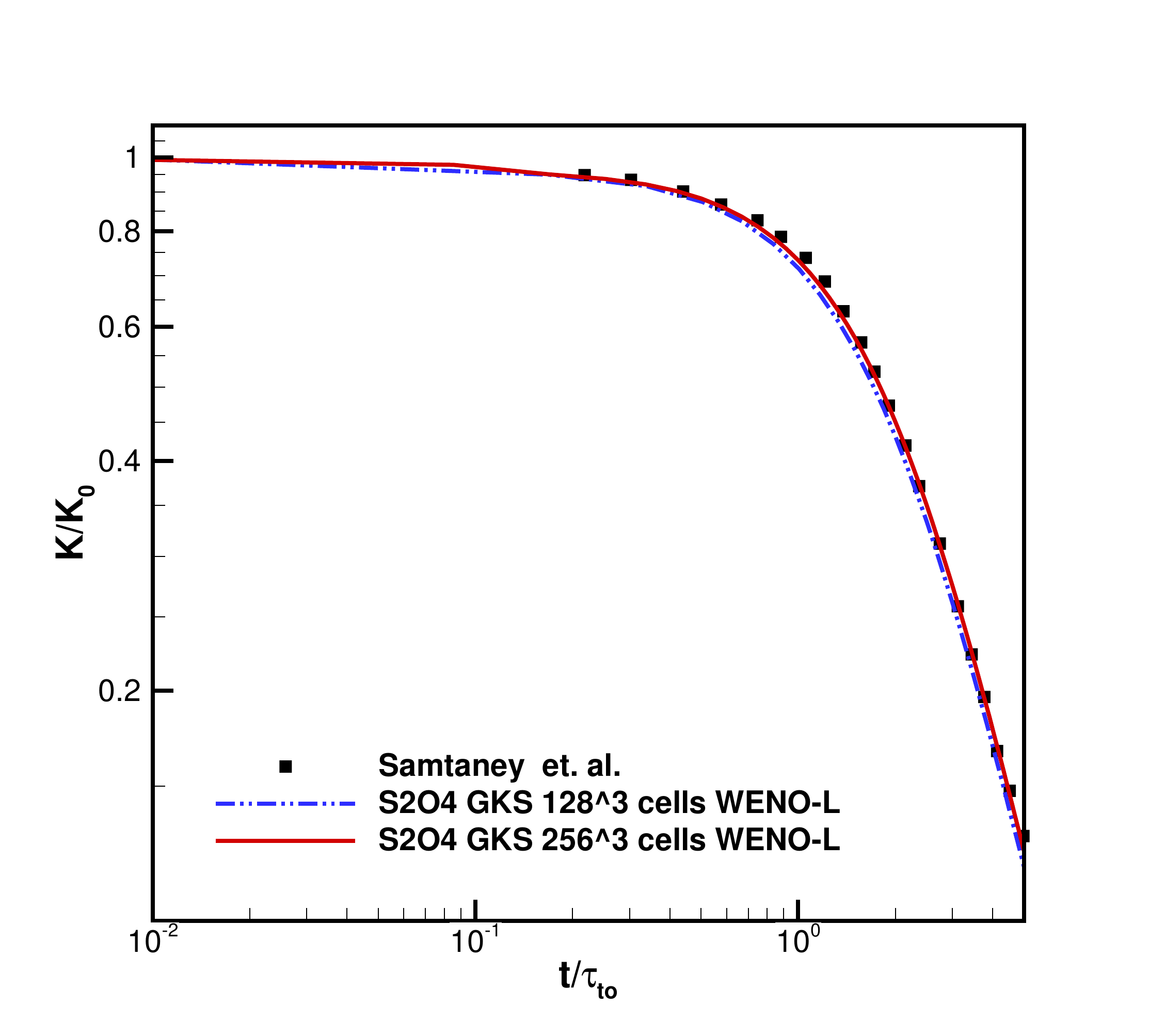}
\caption{\label{Ma03} Time history of $\rho_{rms}/Ma_t^2$ and $K/K_0$ for the isotropic compressible turbulence in nonlinear subsonic regime with $Re_{\lambda}=72$ and $Ma_{t}=0.3$.}
\end{figure}

\subsection{Code validation}
To validate performance of HGKS, the near incompressible isotropic
turbulence with $Ma_{t} = 0.1$ and the compressible isotropic
turbulence in nonlinear subsonic regime with $Ma_{t} = 0.3$ are
tested firstly. In these two cases with low turbulent Mach number,
the flow fields are smooth without strong shocklets. To improve the
resolution of simulation, the WENO scheme with linear weights
denoted as WENO-L is adopted. The uniform grids with $128^3$ and
$256^3$ cells are used. The time history of normalized
root-mean-square density fluctuation $\rho_{rms}/Ma_{t}^2$ and
normalized turbulent kinetic energy $K/K_0$ with respect to
$t/\tau_{t_0}$ are given in Fig.\ref{Ma01} and Fig.\ref{Ma03}. For
these isotropic turbulent flows with low turbulent Mach number, the
WENO-L can well resolve the flow structures. Numerical results of
current scheme agree well with the reference data in
\cite{samtaney2001direct}. Because of the lower dissipation of
WENO-L scheme, the convergent solutions can be provided by a
uniform $128^3$ grid points.

\begin{figure}[!h]
\centering
\includegraphics[width=0.49\textwidth]{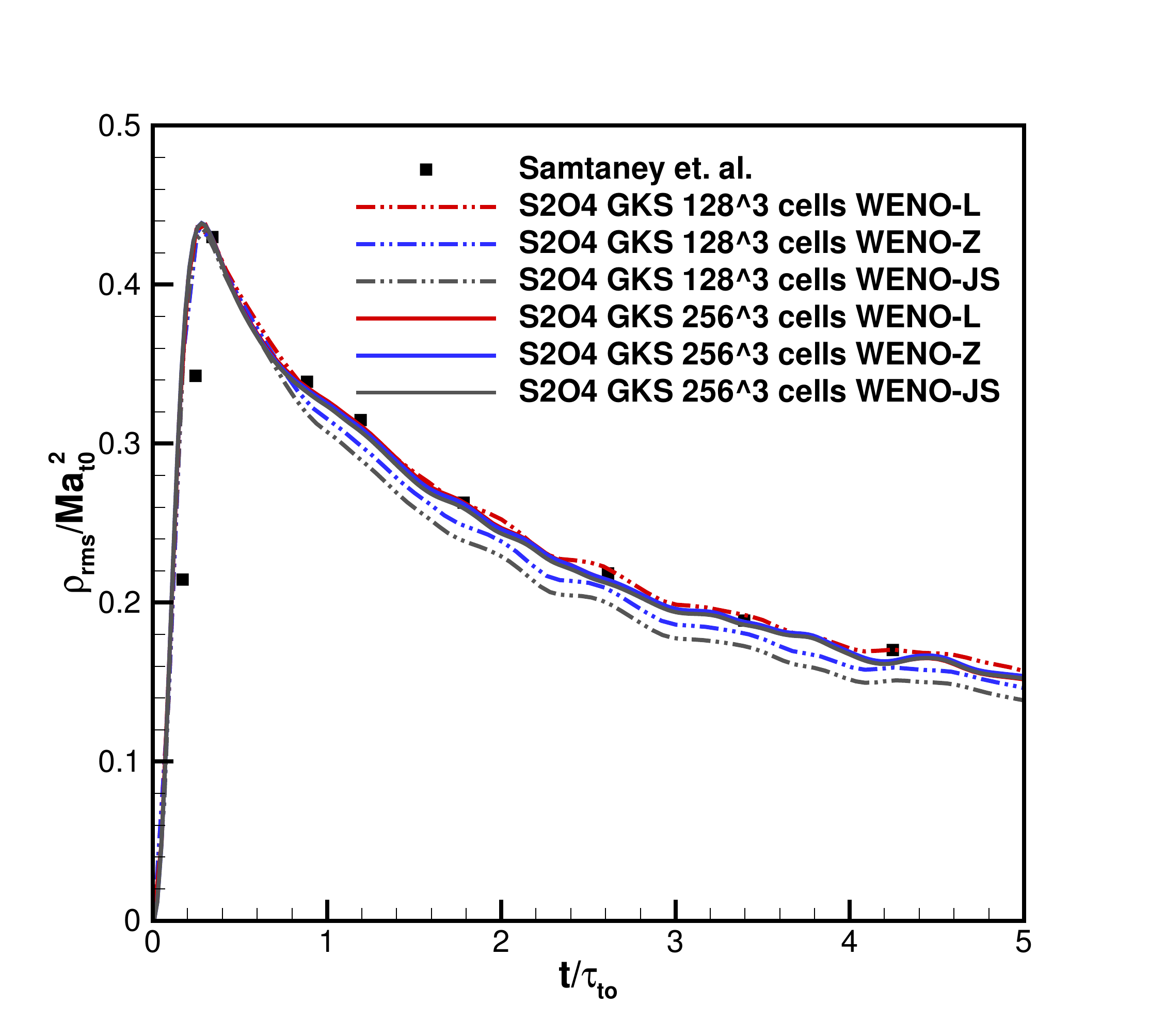}
\includegraphics[width=0.49\textwidth]{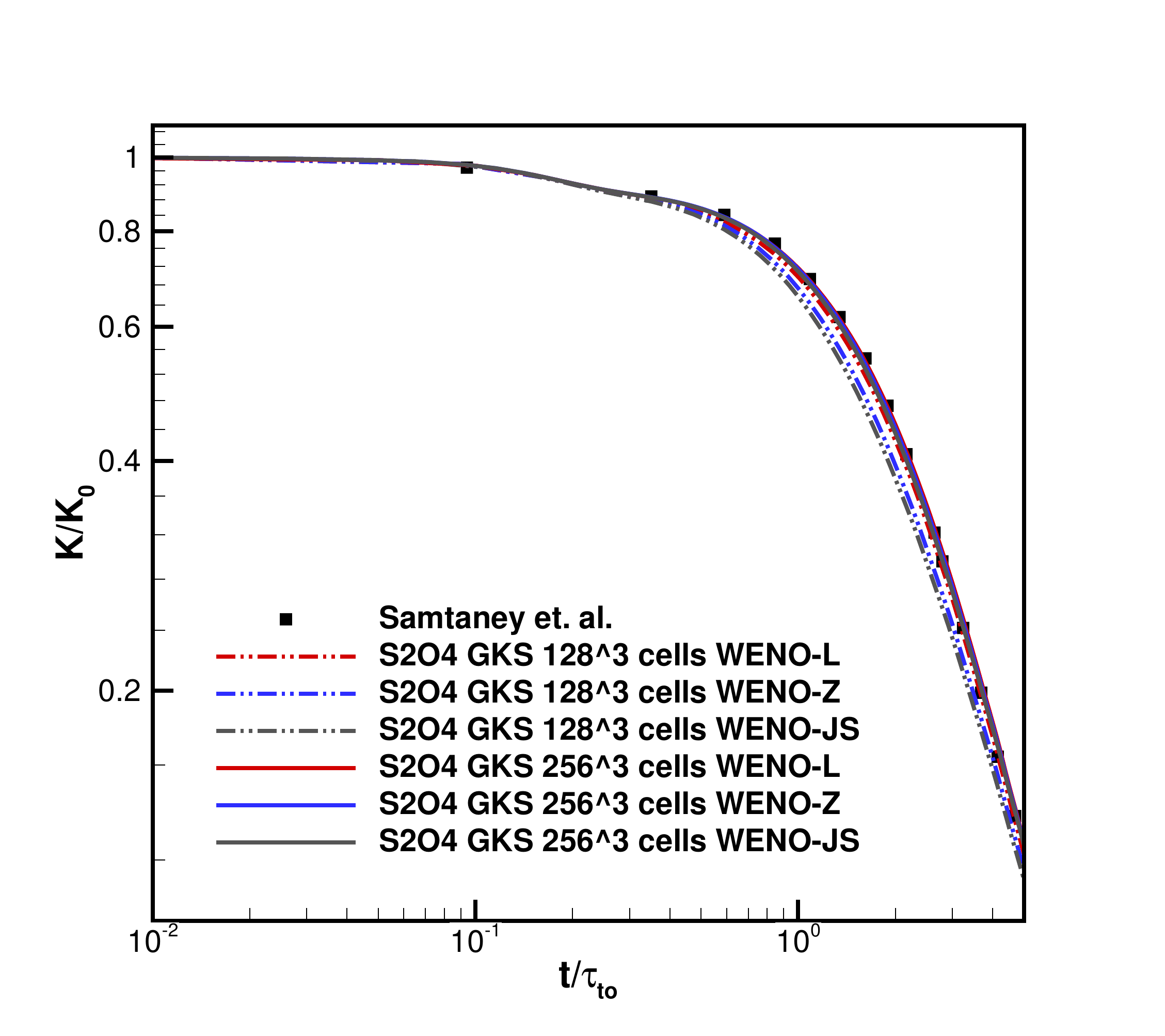}
\caption{\label{ma05weno} Time history of $\rho_{rms}/Ma_{t}^2$ and
$K/K_0$ for the isotropic compressible turbulence with
$Re_{\lambda}=72$ and $Ma_{t}=0.5$ for WENO-L, WENO-JS and WENO-Z
schemes.}
\end{figure}

However, with the increase of turbulent Mach number, the
eddy-shocklets appear in the flow fields and the WENO-L scheme blows
up at $Ma_{t} > 0.5$. Hence, the WENO scheme with nonlinear
weights have to be used to capture the discontinuities when simulating
high turbulent Mach number flows. Before we study the compressible isotropic turbulence in supersonic regime, 
it is legitimate to
study the behavior of high-order GKS with different WENO schemes.
The decaying isotropic compressible turbulence with $Re_{\lambda} =
72$ and $Ma_{t} = 0.5$ is used to test the performance of three
widely used WENO schemes, i.e. WENO-L, WENO-JS and WENO-Z schemes.
The uniform grids with $128^3$ and $256^3$ cells are used as well.
The time history of normalized root-mean-square density fluctuation
$\rho_{rms}/Ma_t^2$ and normalized turbulent kinetic energy $K/K_0$
with respect to $t/\tau_{t_0}$ for three WENO schemes are given in
Fig.\ref{ma05weno}. The convergent solutions can be provided by the
WENO-L scheme with $128^3$ uniform grids, while WENO-JS scheme and
WENO-Z scheme are more dissipative than WENO-L scheme. More
specifically, the WENO-JS is more dissipative than WENO-Z scheme.
According to the numerical tests, the WENO-Z is almost as robust as
WENO-JS for the isotropic compressible turbulence. Considering the
robustness and dissipative behavior, the WENO-Z scheme will be used
in the following simulations.

\begin{table}[!h]
\begin{center}
\def\temptablewidth{0.75\textwidth}
{\rule{\temptablewidth}{1.0pt}}
\begin{tabular*}{\temptablewidth}{@{\extracolsep{\fill}}c|cccc}
CFL number  & $\text{d}t_{ini}/\tau_0$   & $\text{d}t_{end}/\tau_0$    & $\text{d}t_{ini}/\tau_{t_0}$  & $\text{d}t_{end}/\tau_{t_0}$ \\
\hline
0.2         & $9.02/1000$         & $14.00/1000$         & $1.86/1000$           & $2.89/1000$ \\
0.4         & $18.05/1000$        & $28.03/1000$         & $3.72/1000$           & $5.78/1000$  \\
0.6         & $27.08/1000$        & $42.06/1000$         & $5.58/1000$           & $8.66/1000$  \\
\end{tabular*}
{\rule{\temptablewidth}{1.0pt}}
\end{center}
\vspace{-5mm} \caption{\label{ma05_time_table} Different CFL number
for time convergence study.}
\end{table}

\begin{figure}[!h]
\centering
\includegraphics[width=0.49\textwidth]{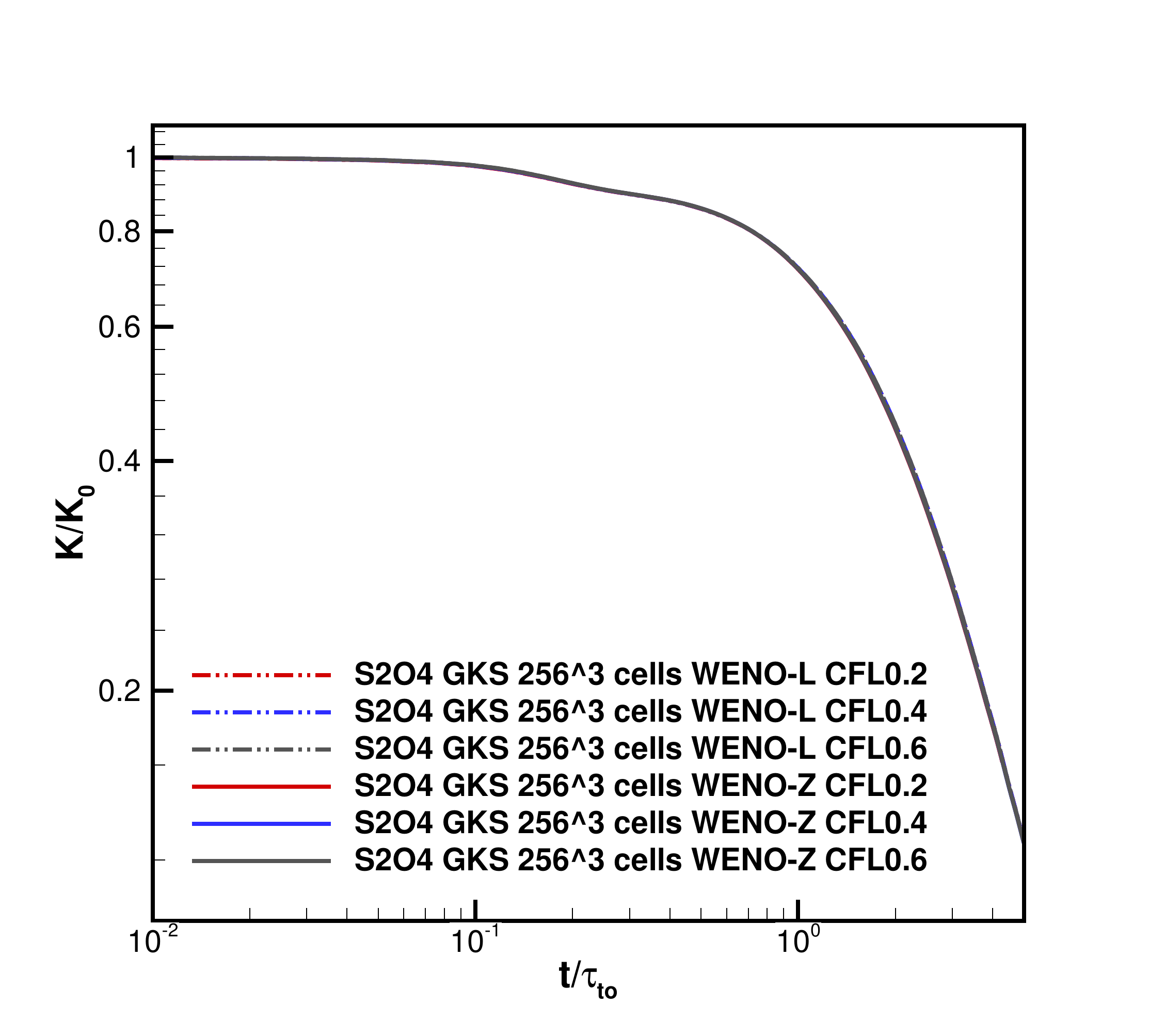}
\includegraphics[width=0.49\textwidth]{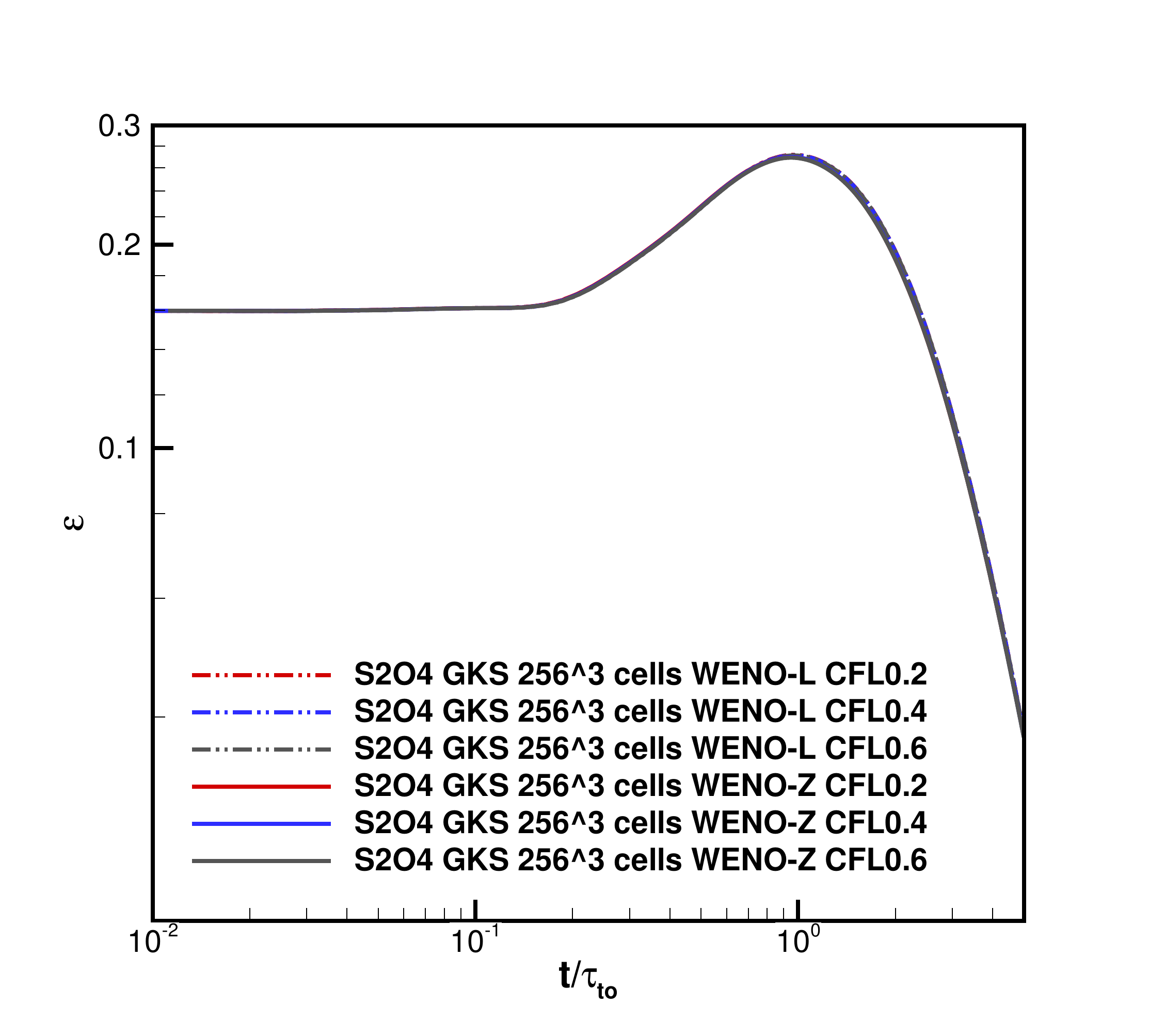}
\caption{\label{ma05_time_convergence} Time convergence study: Time
history of $K/K_0$ and $\varepsilon$ for the isotropic compressible
turbulence with $Re_{\lambda}= 72$ and $Ma_{t} = 0.5$ with CFL
number $0.2$, $0.4$ and $0.6$.}
\end{figure}

The time convergence is studied as well and the same isotropic
compressible turbulence at $Re_{\lambda} = 72$ and $Ma_{t}=0.5$ is
used. In order to resolve and capture the desired physics in the
high turbulent Mach number regime, several strategies for the choice
of time step are provided in previous studies. Time step is set as
$\tau_{t_0}/1000$ \cite{wang2010hybrid} in hybrid scheme, which is
very expansive when implementing DNS on isotropic compressible
turbulence even with moderate Taylor microscale Reynolds number. 
In the DNS using high-resolution modified-WENO GKS
\cite{kumar2013weno}, the maximum CFL number can get up to $0.8$.
Thus, it is legitimate to study the criterion for time step of
current fourth-order GKS. Time convergence study with different CFL
number are presented in Table \ref{ma05_time_table}, where
$\text{d}t_{ini}$ and $\text{d}t_{end}$ represent the time step for
the initial step and ending step, respectively. In this simulation,
the ending step is defined at the moment of $t/\tau_{t_0}=5$. The
time history of normalized $K/K_0$ and the ensemble total
dissipation rate $\varepsilon$ with CFL number $CFL=0.2, 0.4$ and
$0.6$ on uniform grids with $256^3$ cells are shown in Fig
\ref{ma05_time_convergence}. In current paper, the velocity
gradients for Eq.(\ref{dkdt}) are computed by first-order upwind
scheme.  As the consistent results are obtained using different CFL
number with WENO-L and WENO-Z schemes, time convergent solution can
be obtained with CFL number $CFL=0.6$, which is in agreement with
modified-WENO GKS \cite{kumar2013weno}. For this case, the initial
Kolmogorov time scale $\tau_0$ and the initial large-eddy turnover
time $\tau_{t_0}$ can be determined by Eq.\eqref{initial_def}.
According to Table.\ref{ma05_time_table}, the time step can be set
as large as $t_{ini}/\tau_{t_0} = 5.58/1000$. Meanwhile, the time
step can well resolve the smallest timescale $t_{ini}/\tau_0 =
27.08/1000$.

\begin{figure}[!h]
\centering
\includegraphics[width=0.49\textwidth]{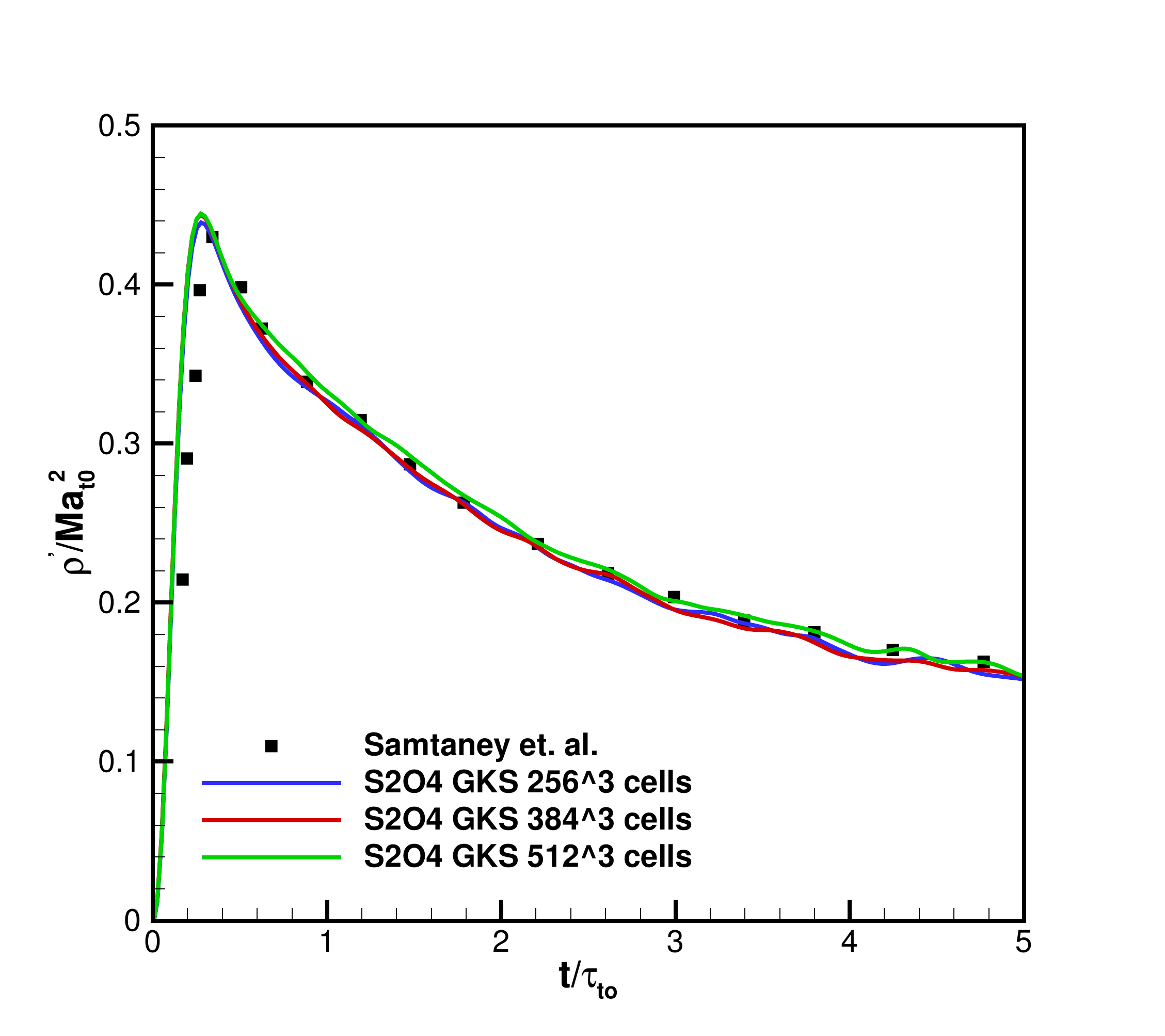}
\includegraphics[width=0.49\textwidth]{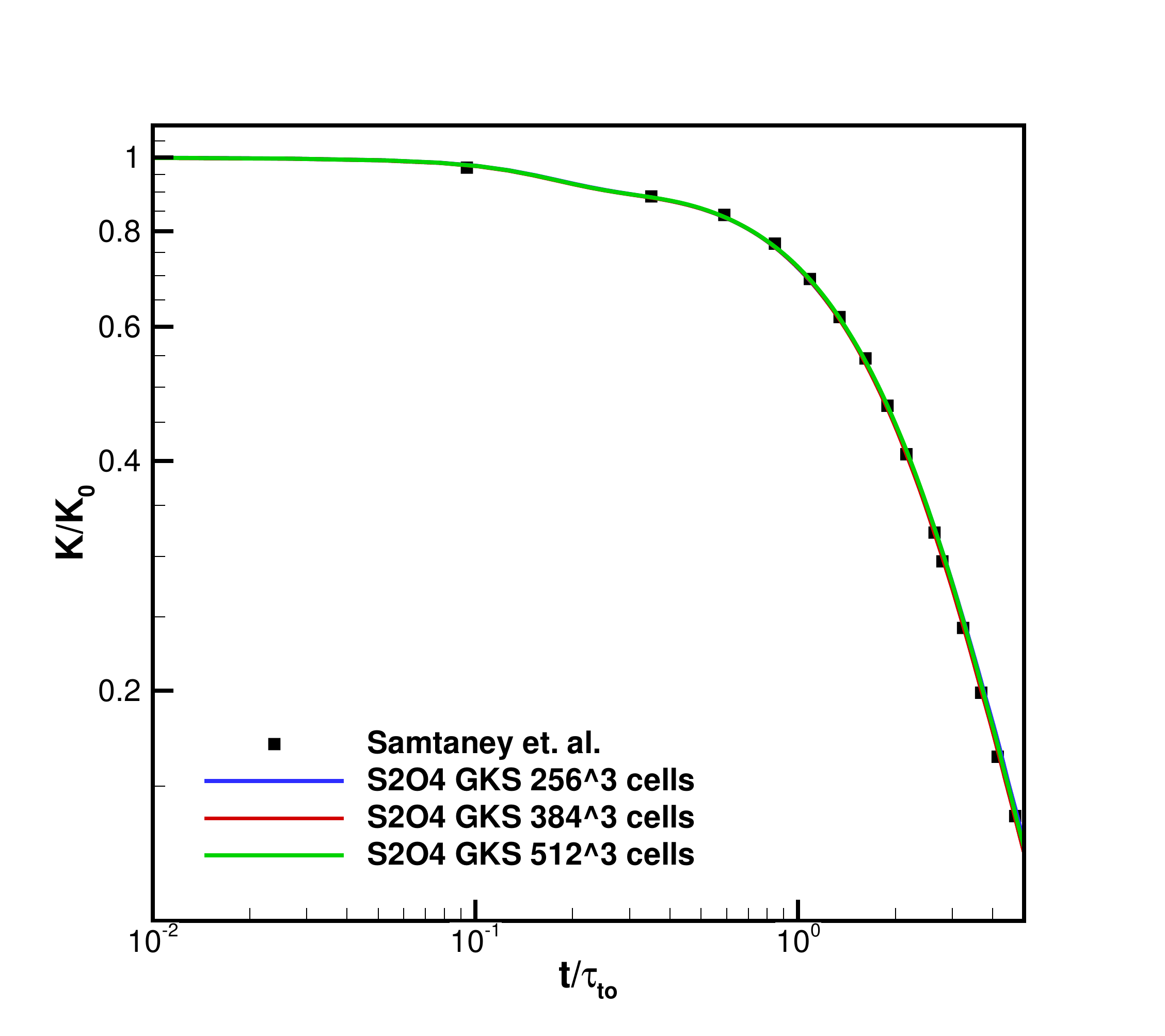}
\includegraphics[width=0.49\textwidth]{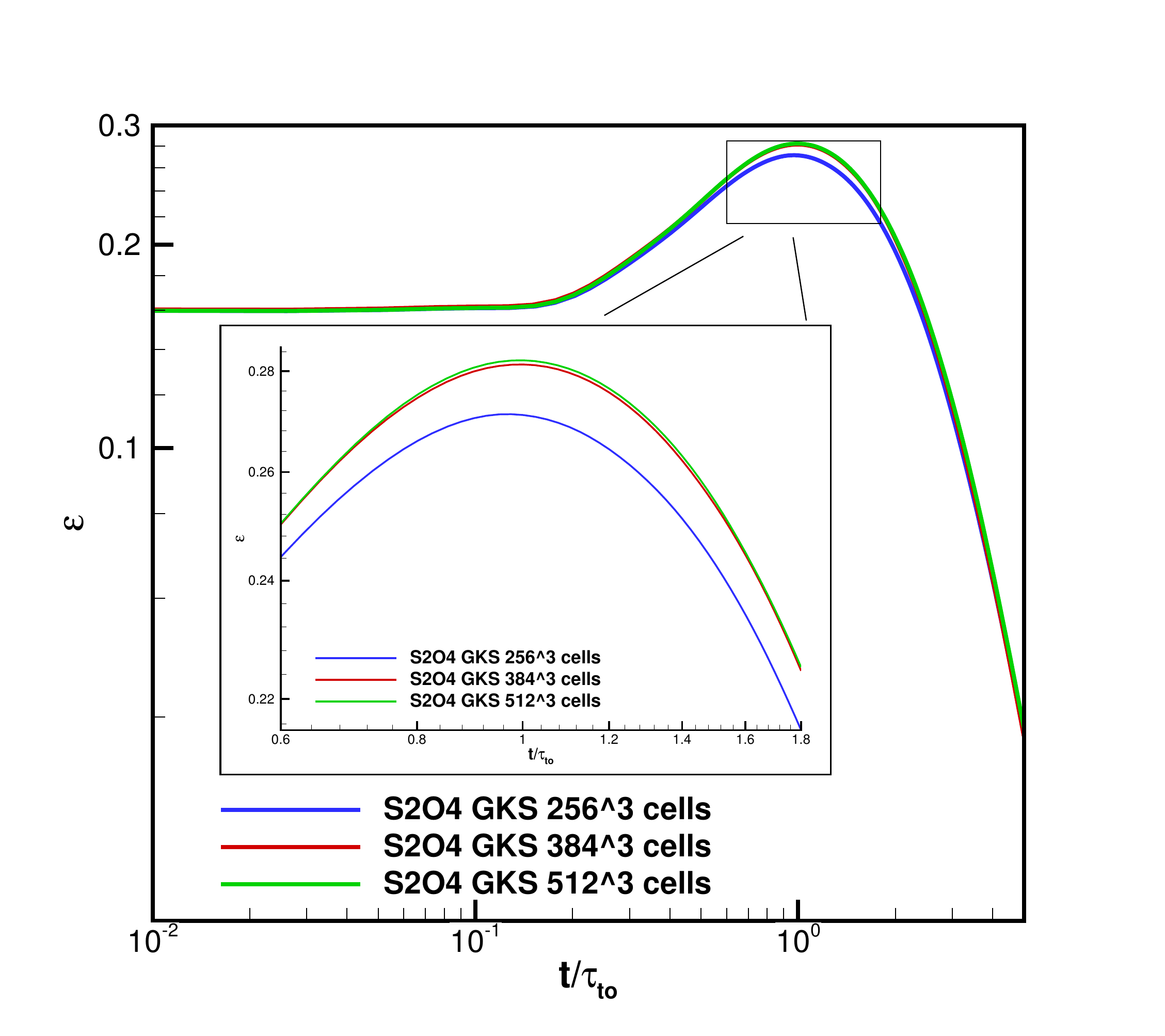}
\includegraphics[width=0.49\textwidth]{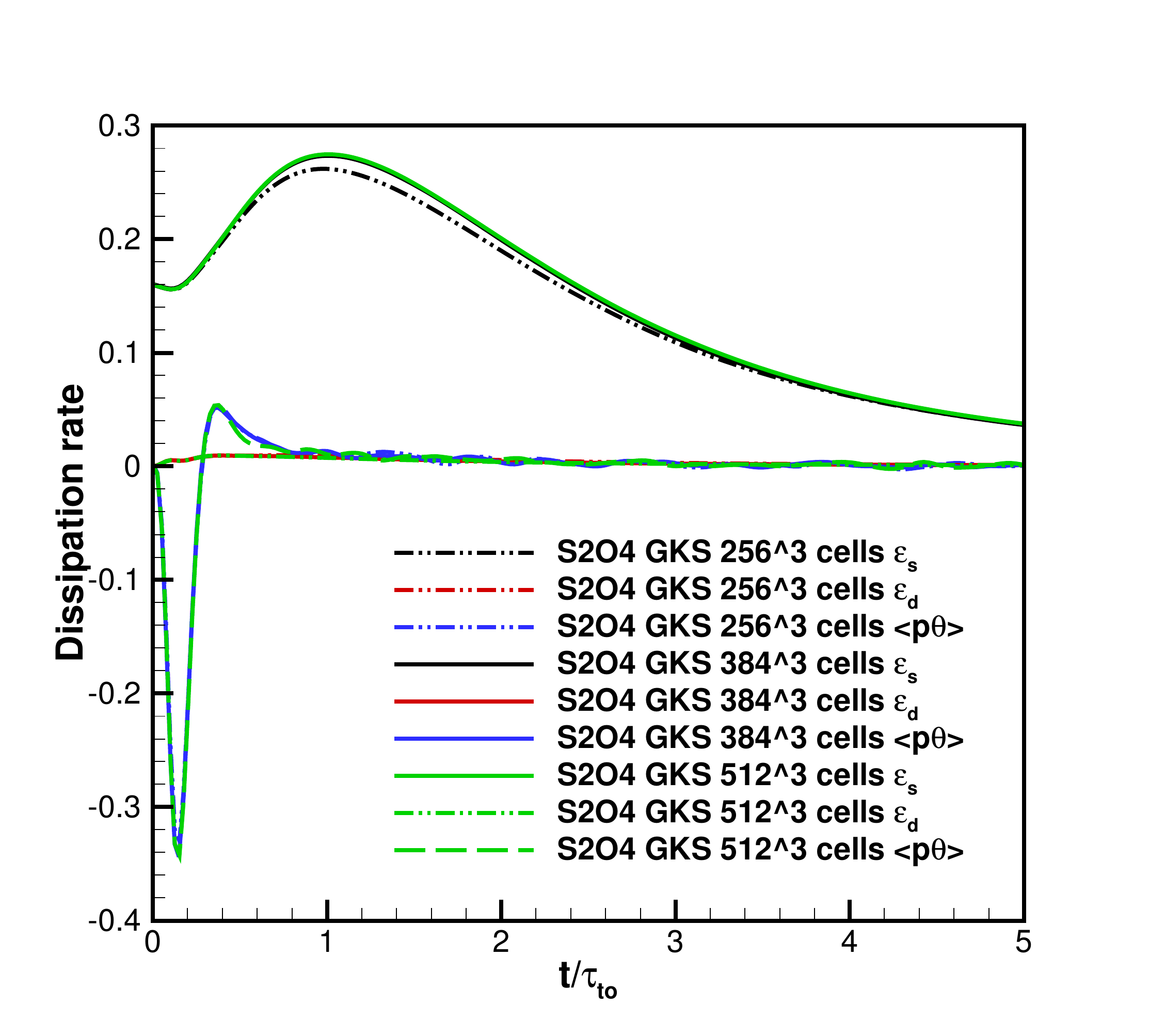}
\caption{\label{ma05_grid_convergence} Grid convergence study: Time
history of $\rho_{rms}/Ma_{t}^2$, $K/K_0$, $\varepsilon$,
$\varepsilon_d$, $\varepsilon_s$ and $\left\langle p \theta
\right\rangle$ for isotropic compressible turbulent with
$Re_{\lambda} = 72$ and $Ma_{t} = 0.5$ on uniform grids with
$256^3$, $384^3$ and $512^3$ cells.}
\end{figure}

\begin{table}[!htp]
\begin{center}
\def\temptablewidth{0.5\textwidth}
{\rule{\temptablewidth}{1.0pt}}
\begin{tabular*}{\temptablewidth}{@{\extracolsep{\fill}}c|ccc}
Grid size    &$\Delta/\lambda_0$  &$\Delta/\eta_0$          &$\kappa_{max}\eta_0$ \\
\hline
$256^3$      & $8.162$            & $1.639$                 & $1.806$               \\
$384^3$      & $5.441$            & $1.093$                 & $2.710$               \\
$512^3$      & $4.081$            & $0.819$                 & $3.613$               \\
\end{tabular*}
{\rule{\temptablewidth}{1.0pt}}
\end{center}
\vspace{-5mm} \caption{\label{ma05gridtable} Grid size and characteristic length scales for grid convergence study.}
\end{table}

Grid convergence study is also required
\cite{wang2010hybrid, wang2016comparison} to conclude the criterion
for space resolution when using the fourth-order GKS as a DNS tool.
Three different uniform grids with $256^3$, $384^3$ and $512^3$
cells and characteristic length scales are demonstrated in Table
\ref{ma05gridtable}, where $\lambda_0$ is the initial mean free path
approximated by $\mu_0 = 1/3 \rho_0 c_0 \lambda_0$
\cite{xu2015direct}, $\Delta$ is the uniform grid size in each
direction, $\eta_0$ is the initial Kolmogorov length scale as in
Eq.(\ref{initial_def}), $\kappa_{max} = \sqrt{2} \kappa_0 N/3$ is
the maximum resolved number wave number
\cite{eswaran1988examination}, $\kappa_0 = 8$ as
Eq.(\ref{initial_spectrum}) and $N$ is the number of grid points in
each Cartesian direction. According to Table \ref{ma05gridtable},
the Kolmogorov length scale is almost   $5$ times larger than the
mean free path, and each grid always contains several mean free path
even for the finest grids with $512^3$ cells. This provides the
intuitive evidence for controversial issue that smallest eddies in
turbulence may still within the framework of continuum mechanics
assumption. The behavior of normalized $\rho_{rms}/Ma_{t}^2$,
$K/K_0$ and turbulent kinetic energy budget defined in
Eq.\eqref{dkdt} are presented in Fig \ref{ma05_grid_convergence}.
The key statistical quantities on uniform grids with $384^3$ cells
coincide with those on uniform grids with $512^3$ cells. It can be
concluded that the minimum spatial resolution parameter
$\kappa_{max} \eta_0 \ge 2.71$ is adequate for resolving the
isotropic compressible turbulence of HGKS. This criterion is similar
to that in the hybrid scheme \cite{wang2010hybrid}, which has been applied in
isotropic compressible turbulence successfully. According to the
simulations above, the criterion of spatial and temporal resolution
for fourth-order GKS based on WENO-Z reconstruction is obtained,
which will be adopted for the simulation of high turbulent Mach
number isotropic compressible turbulence up to supersonic regime.

\begin{figure}[!h]
\centering
\includegraphics[width=0.49\textwidth]{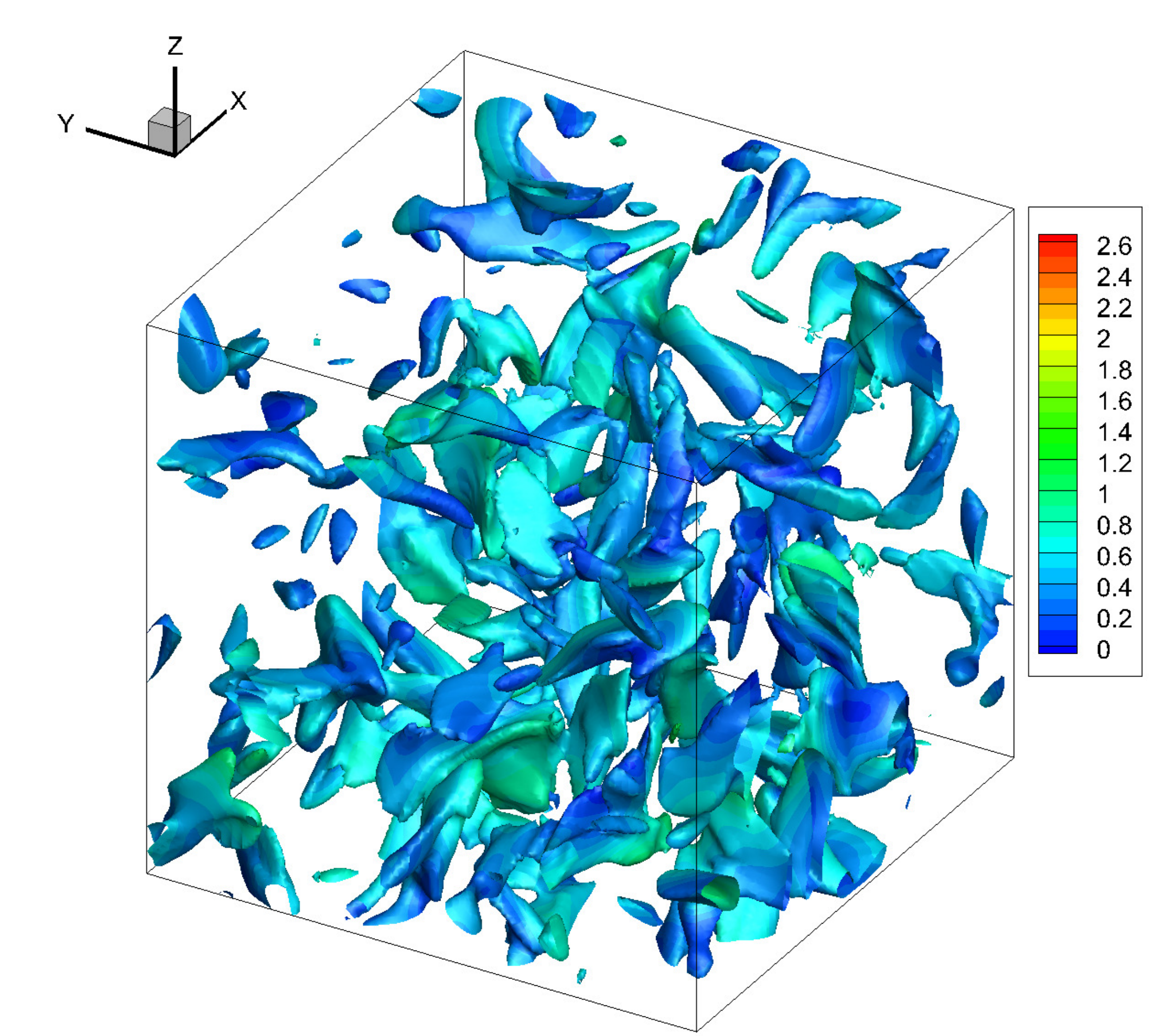}
\includegraphics[width=0.49\textwidth]{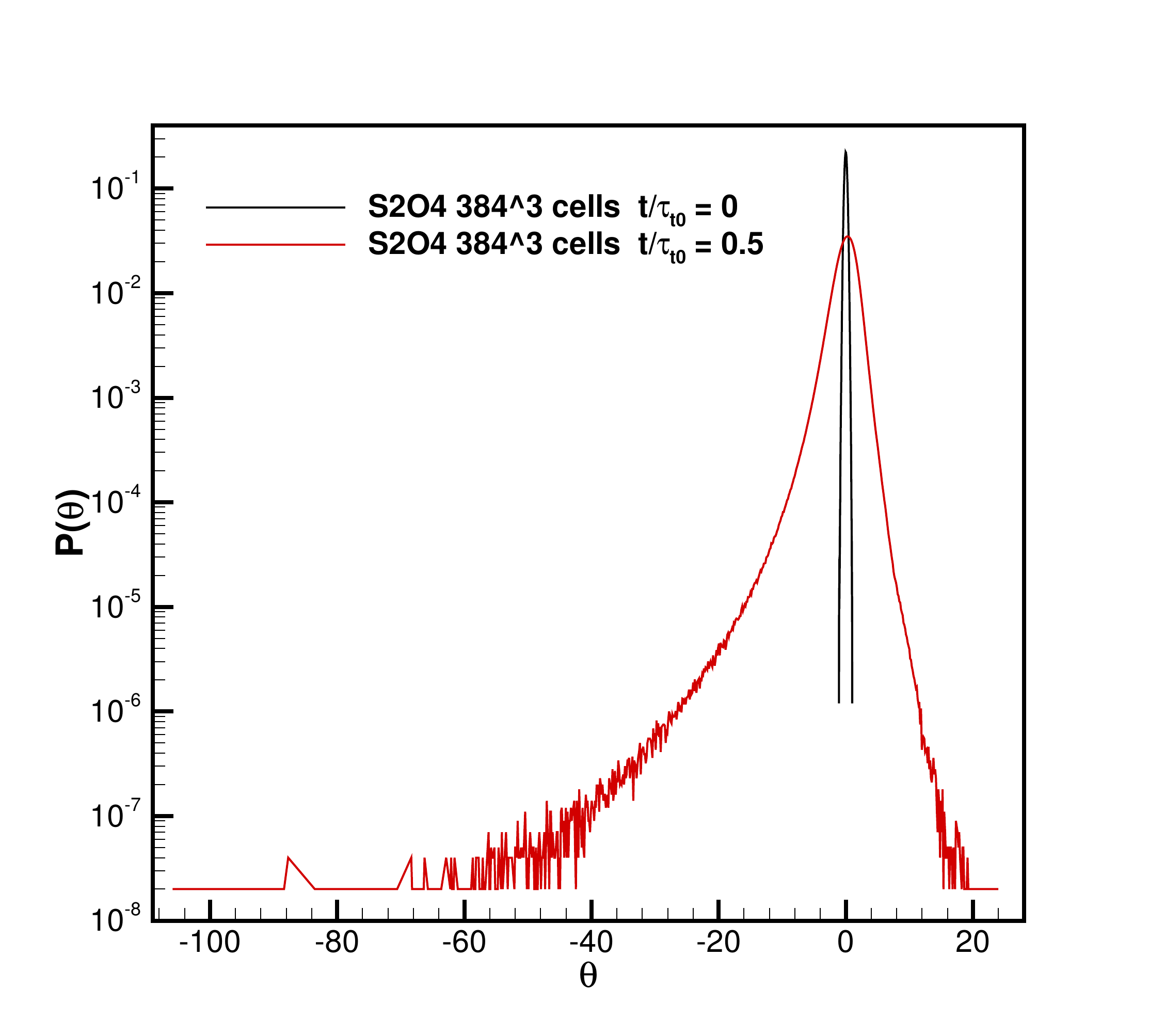}
\caption{\label{ma05_pmat}Iso-surface of the second invariant of
velocity gradient tensor $Q = 25$ and PDF of dilation $\theta$ with
$Re_{\lambda} = 72$ and $Ma_{t} = 0.5$ on uniform grids with $384^3$
cells at $t/\tau_{t_0}= 0.5$. }
\end{figure}

Iso-surface of the second invariant of velocity gradient tensor $Q =
25$ and PDF of dilation $\theta$ for
uniform grids with $384^3$ cells at $t/\tau_{t_0} = 0.5$ are shown
in Fig \ref{ma05_pmat}. Iso-surface is colored by local turbulent
Mach number in a $128^3$ sub-domain covering $(105 \eta_0)^3$ (1/27
of the whole domain), where the sub-domain is located at the center
of the full domain. The local turbulent Mach number concentrates on
the near region of $Ma_t = 0.4$, which is much smaller than the
latter simulation with high turbulent Mach numbers. All PDFs of
dilation in this paper are obtained by dividing the dilation range
into $1000$ equivalent intervals, the velocity gradients for local
dilation value $\theta$ are computed by the second-order central
difference. At the initial stage, the symmetric dilation range is
$[-1.08, 1.08]$, while the skewed dilation range with  minimum and
maximum values $-105.8$ and $23.9$ at $t/\tau_{t_0} = 0.5$ appears.
This quite wide range of dilation is an intrinsic property for
isotropic compressible turbulence which means the strong compression
and expansion regions exist in the flow field.

\begin{table}[!h]
\begin{center}
\def\temptablewidth{0.9\textwidth}
{\rule{\temptablewidth}{1.0pt}}
\begin{tabular*}{\temptablewidth}{@{\extracolsep{\fill}}c|cccccc}
Test                     &$R_0$    &$R_1$        &$R_2$       &$R_3$       &$R_4$       &$R_5$  \\
\hline
$Ma_{t}$               &0.5   &0.8       &0.9      &1.0      &1.1      &1.2 \\
$\text{d}t_{ini}/\tau_{t_0}$    &$5.58/1000$      &$4.61/1000$    &$4.89/1000$     &$3.76/1000$ &$3.85/1000$ &$3.92/1000$ \\
$\left\langle \theta \right\rangle^{\ast}$  &2.12   &3.39     &3.87     &4.28      &4.65    &4.90 \\
\end{tabular*}
{\rule{\temptablewidth}{1.0pt}}
\end{center}
\vspace{-5mm} \caption{\label{re72gridtable} Isotropic compressible turbulence with different high turbulent Mach number.}
\end{table}

\subsection{Turbulent Mach number effect}
\begin{figure}[!h]
\centering
\includegraphics[width=0.49\textwidth]{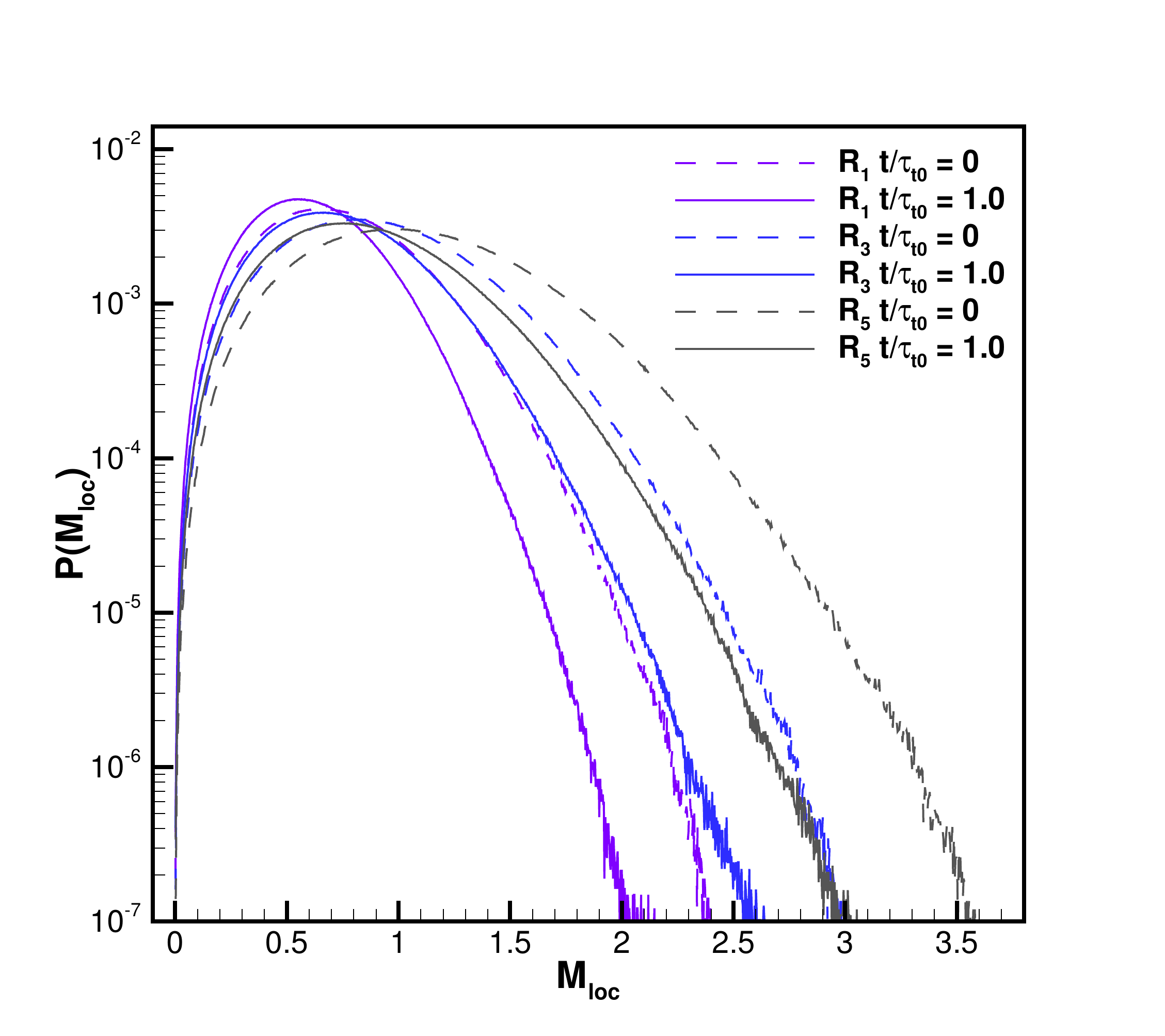}
\includegraphics[width=0.49\textwidth]{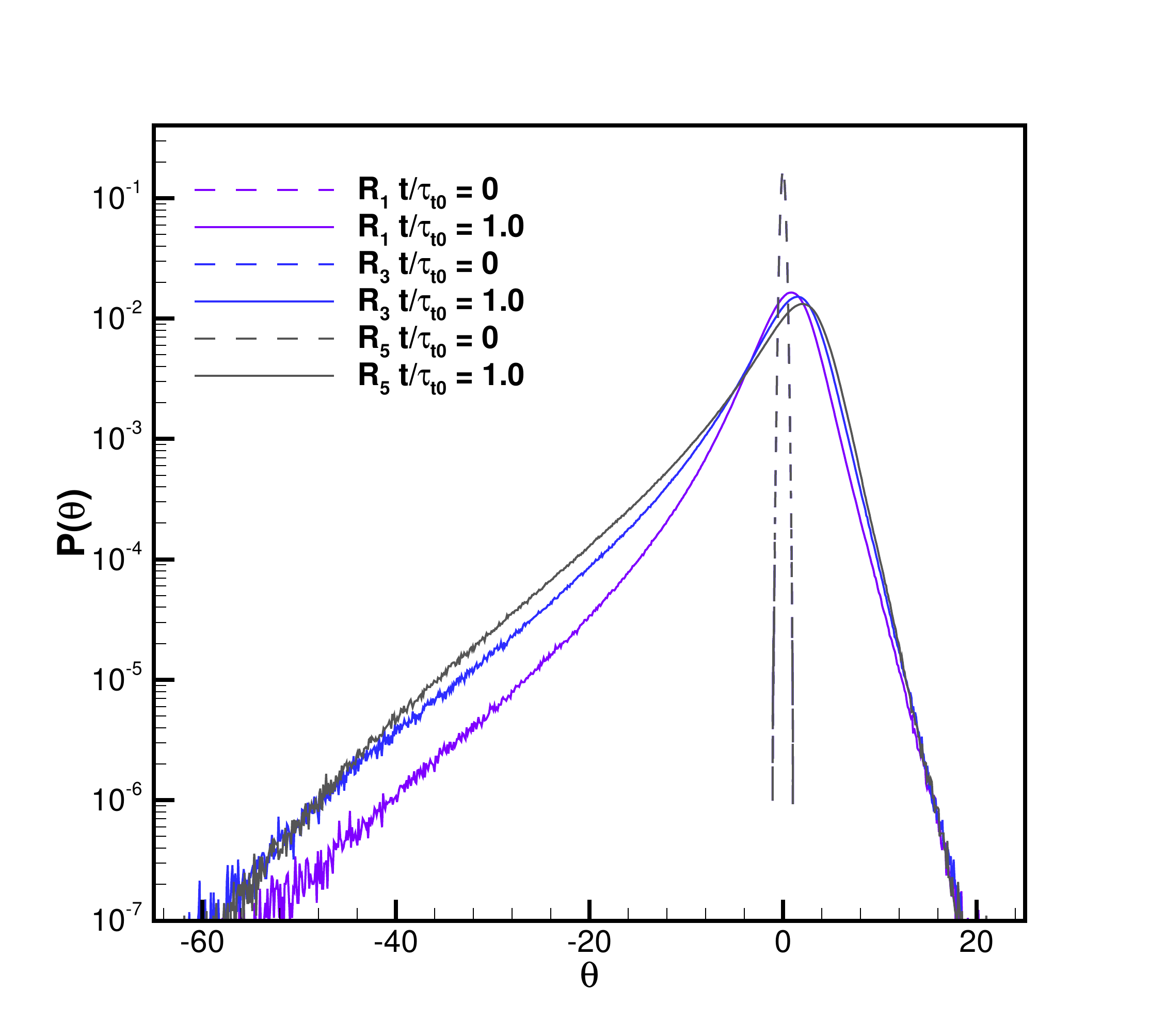}
\caption{\label{pdf_mloc_theta_ma} PDF of local turbulent Mach
number and root-mean-square dilation at $t/\tau_{t_0} = 0$ and
$t/\tau_{t_0} = 1.0$ for $R_1$, $R_3$ and $R_5$.}
\end{figure}

In this section, DNS of isotropic compressible turbulence from high
subsonic regime to supersonic regime with moderate Taylor microscale
Reynolds number are tested. The effect of compressibility on
dynamics and structures of isotropic compressible turbulence in
moderate subsonic regime ($Ma_{t} \leq 0.8$) has been studied
previously in \cite{sarkar1991analysis, pirozzoli2004direct}. The
statistical properties and dynamics of forced supersonic regime
$Ma_t \approx 1.0$ have been studied systematically in
\cite{wang2012scaling, wang2017scaling, wang2017shocklet,
xie2018modified}. The numerical tests $R1-R5$ given in
Table.\ref{re72gridtable} go beyond previous study up to the maximum
supersonic turbulent Mach number $Ma_t=1.2$ with a fixed Taylor
microscale Reynolds number $Re_{\lambda} = 72$. In the computation,
a uniform grids with $384^3$ cells are used and
$\kappa_{max}\eta_0=2.71$. As shown in Fig.\ref{pdf_mloc_theta_ma},
the PDF of initial local turbulent Mach number deviates from symmetric
distribution and the maximum local turbulent Mach number for $R_1$,
$R_3$ and $R_5$ can be three times higher  than ensemble initial
turbulent Mach number $Ma_{t_0}$. For this decaying isotropic
compressible turbulence, the ensemble turbulence Mach number becomes
smaller monotonically.  After a long decay at $t/\tau_{t_0} = 1.0$,
PDFs of the local turbulent Mach number still show large portion of
flow fields in supersonic state. It means that strong
shocklets randomly distribute in flow fields. These random strong
discontinuities really pose a great challenge for high-order
schemes, which have to well resolve the small scales in smooth
regions as well as capture the shock sharply. The PDFs of dilation
are presented in Fig \ref{pdf_mloc_theta_ma} as well. At the
beginning, all PDFs are symmetric and in a narrow range from $-1.08$
to $1.08$. With the evolving of flows, the systems experience a
sharp increase of dilation. The PDFs range approximately from $-60$
to $20$ at $t/\tau_{t_0} = 1.0$, which means the strong compression
and expansion regions appear in flow fields.  All PDFs show strong
negative tales, which are the most significant flow structures of
isotropic compressible turbulence resulting from the shocklets. In
particular, the proportion of negative tail of supersonic isotropic
turbulence is larger than that of high subsonic regime, which
indicates that shocklets appear in the former case more frequently than that
in the latter one. The root-mean-square dilation $\left\langle \theta
\right\rangle^{\ast}$ at $t/\tau_{t_0} = 1.0$ of these cases are
given in Table \ref{re72gridtable}, and it can be concluded that the
higher initial turbulent Mach number possess a much higher
root-mean-square dilation, i.e. the stronger compressibility effect
of isotropic compressible turbulence in supersonic regime. Focusing
on physical mechanism of isotropic compressible turbulence, DNS on
much higher turbulent Mach number up to $Ma_{t} = 2.2$ and higher
Taylor microscale Reynolds number $Re_{\lambda}=100$ have been
obtained by current scheme, which will be presented in the coming
paper. 

\begin{figure}[!h]
\centering
\includegraphics[width=0.49\textwidth]{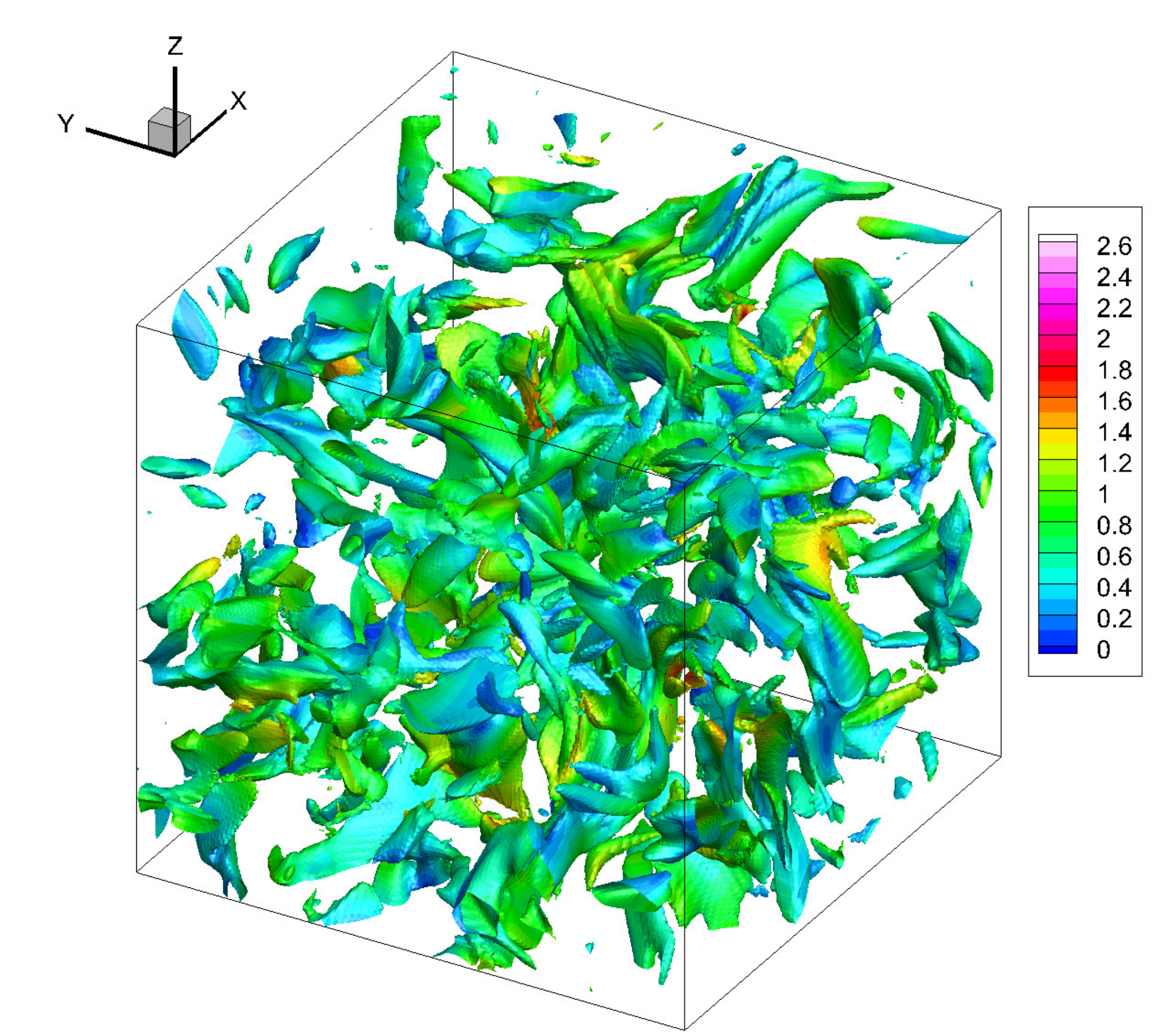}
\includegraphics[width=0.49\textwidth]{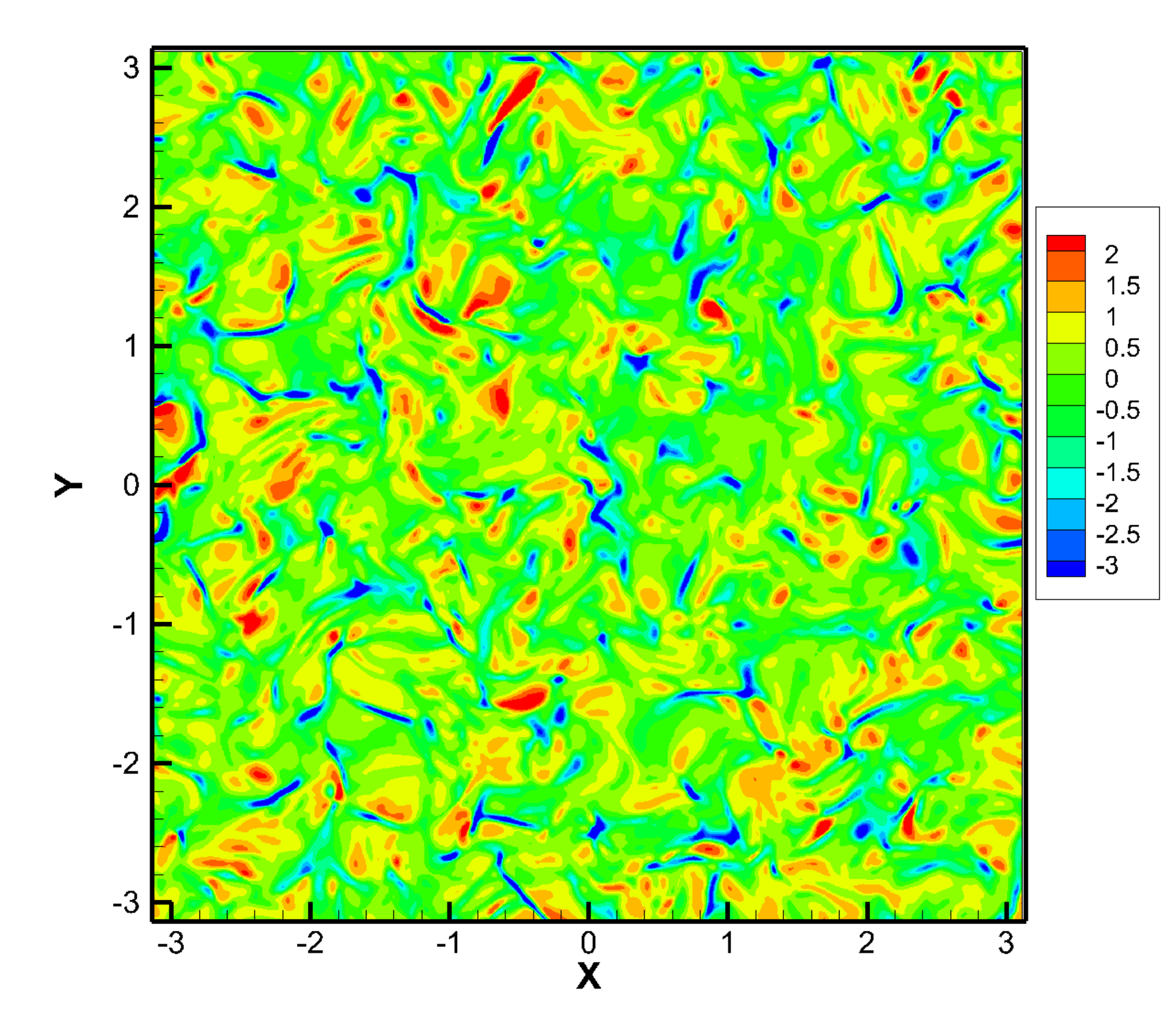}
\caption{\label{Ma08_effect_Q_theta} Iso-surface of the second
invariant of velocity gradient tensor $Q = 25$ and contour of
normalized dilation $\theta/\left\langle \theta
\right\rangle^{\ast}$ on $z= 0$ slice with $Re_{\lambda} =72$ and
$Ma_{t} = 0.8$ at $t/\tau_{t_0} = 1.0$.} \centering
\includegraphics[width=0.49\textwidth]{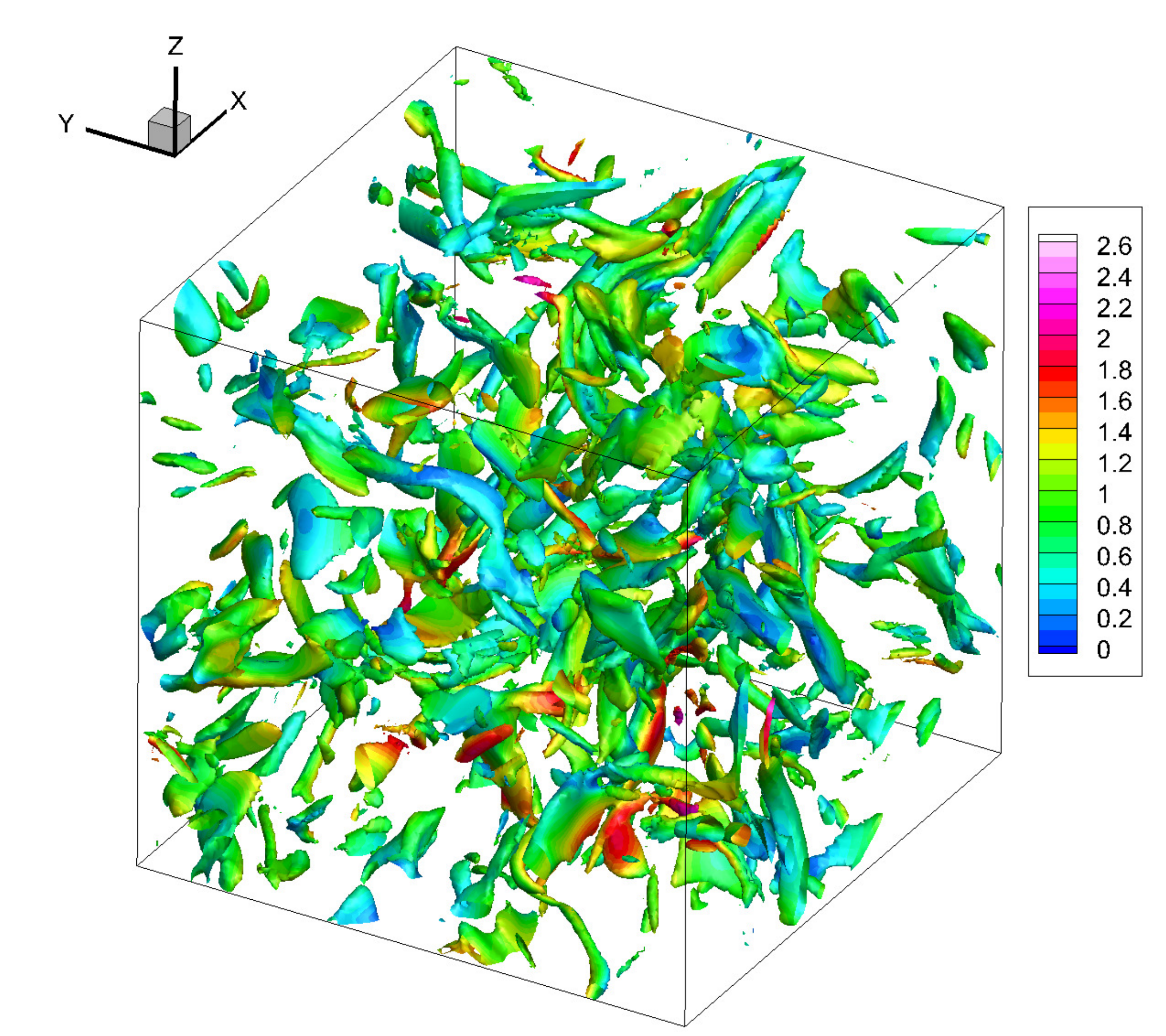}
\includegraphics[width=0.49\textwidth]{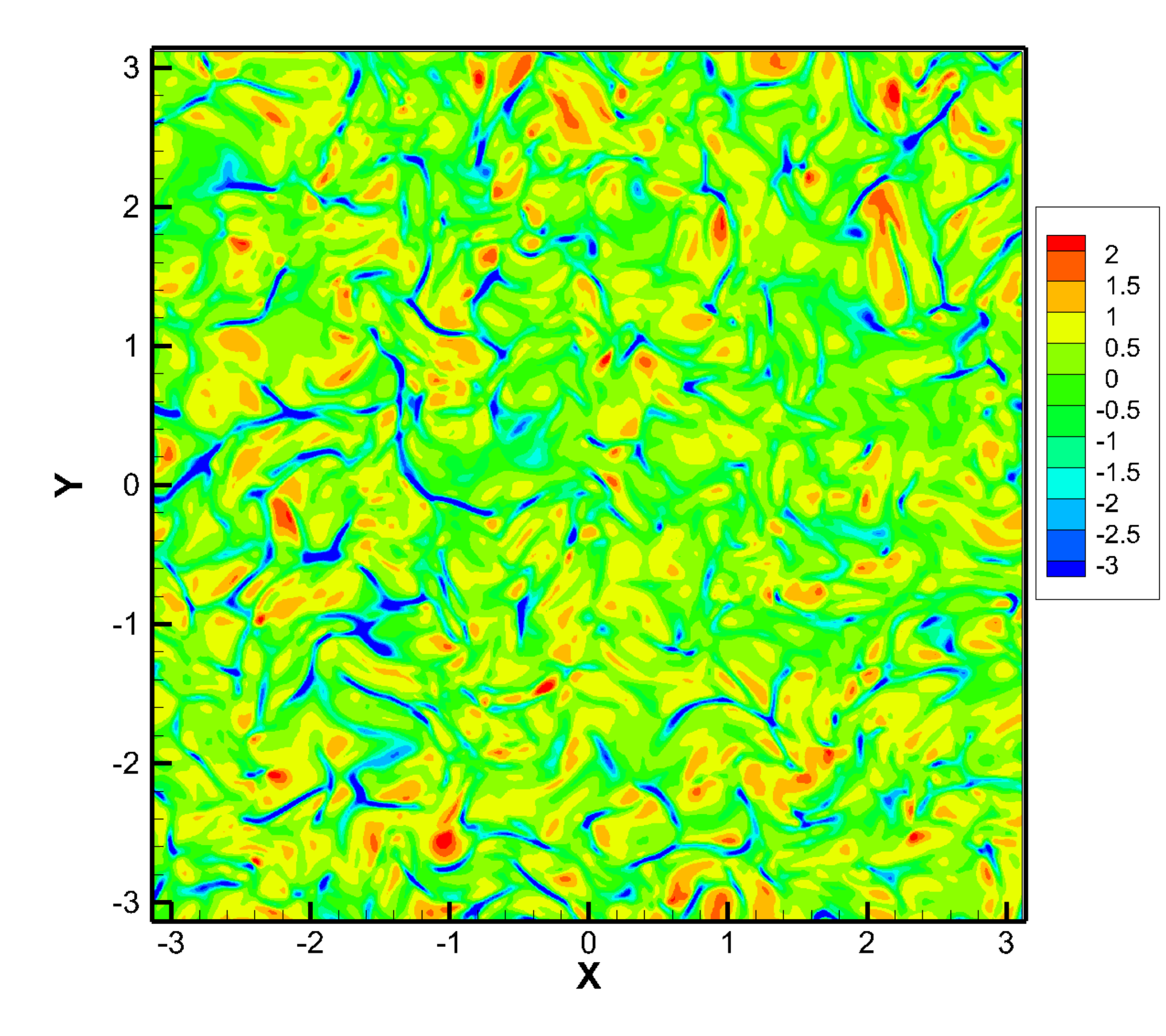}
\caption{\label{Ma12_effect_Q_theta} Iso-surface of the second
invariant of velocity gradient tensor $Q = 25$ and contour of
normalized dilation $\theta/\left\langle \theta
\right\rangle^{\ast}$ on $z= 0$ slice with $Re_{\lambda}=72$ and
$Ma_{t}=1.2$ at $t/\tau_{t_0}=1.0$.  }
\end{figure}

To study the behavior of supersonic isotropic compressible
turbulence further, the iso-surfaces of second invariant of velocity
gradient tensor $Q = 25$ and contours of normalized dilation
$\theta/\left\langle \theta \right\rangle^{\ast}$ on $z=0$ slices
with $Ma_{t}=0.8$ and $Ma_{t}=1.2$ are presented in
Fig.\ref{Ma08_effect_Q_theta} and Fig.\ref{Ma12_effect_Q_theta},
respectively. Iso-surfaces are also colored by the local turbulent
Mach number in a $128^3$ sub-domain covering $(105 \eta_0)^3$, where
the sub-domain is located at the center of the whole domain. A quite
wide range of vortex structure is presented in flow fields for both
cases, and the supersonic isotropic compressible turbulence shows a
much higher local turbulent Mach number region than the subsonic one
after the same decay. Contours of normalized dilation
$\theta/\left\langle \theta \right\rangle ^{\ast}$ shows very
different behavior between the compression motion and expansion
motion. Strong compression regions $\theta/\left\langle \theta
\right\rangle^{\ast} \leq -3$ are usually recognized as shocklets
\cite{samtaney2001direct}. In current study, shocklets behave in the
shape of narrow and long ``ribbon", while high expansion regions
$\theta/\left\langle \theta \right\rangle^{\ast} \geq 2$ are in the
type of localized ``block". In addition, strong compression regions
are close to several regions of high expansion. This behavior is
consistent with the physical intuitive that expansion regions can be
identified just downstream of shock waves \cite{wang2018effect}.
These random distributed shocklets and high expansion region lead to
strong spatial gradient in flow fields. Compared with $R_1$ in
subsonic regime, the supersonic case $R_5$ contains much more
crisp shocklets, which pose much greater challenge for high-order
schemes when implementing DNS for isotropic turbulence in supersonic
regime. In comparison with previous studies, much higher turbulent Mach number can be
simulated by the current scheme, which provide confidence on HGKS for
the study of challenging compressible turbulence problems, such as shock-boundary interaction.

\begin{figure}[!h]
\centering
\includegraphics[width=0.45\textwidth]{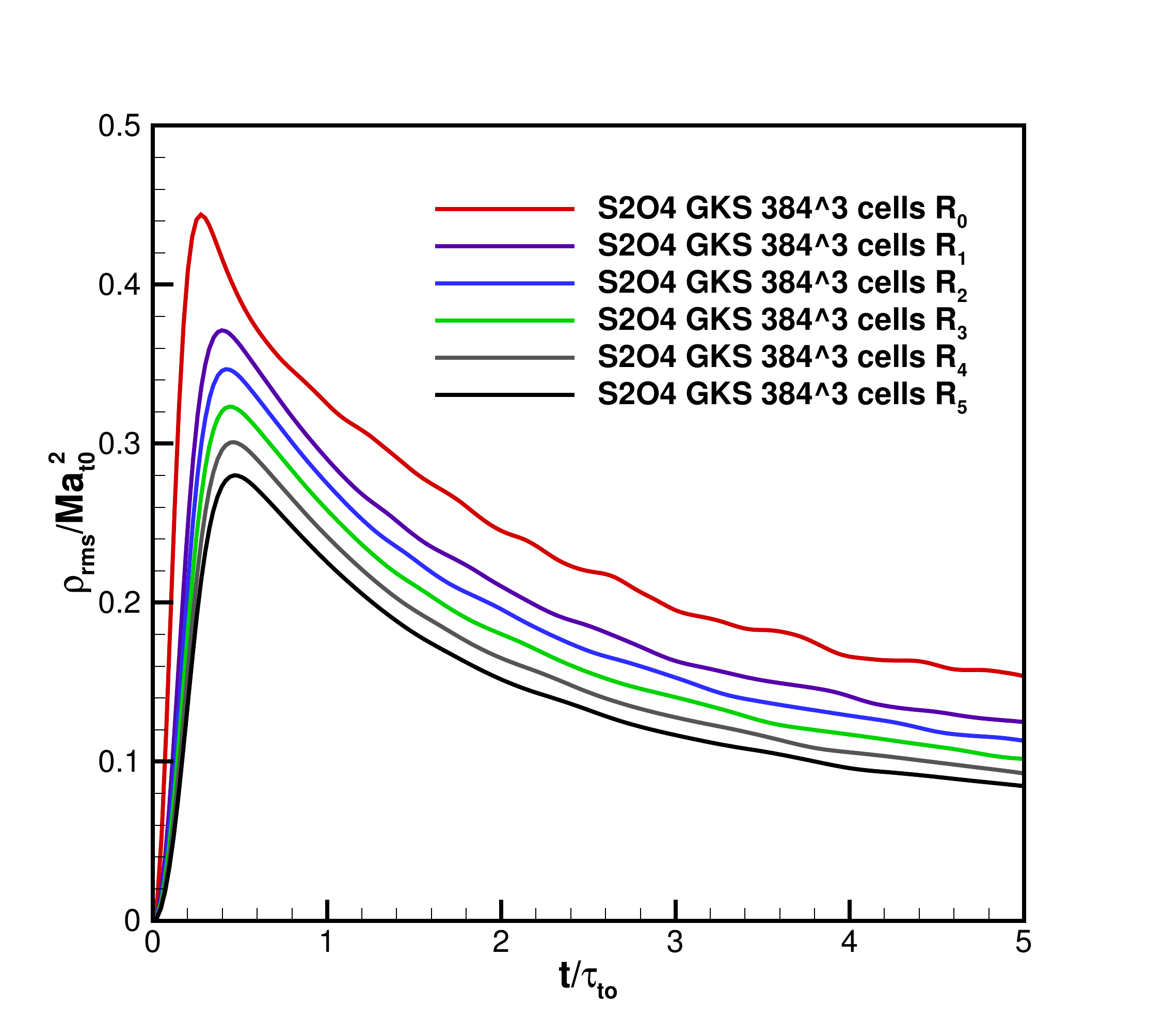}
\includegraphics[width=0.45\textwidth]{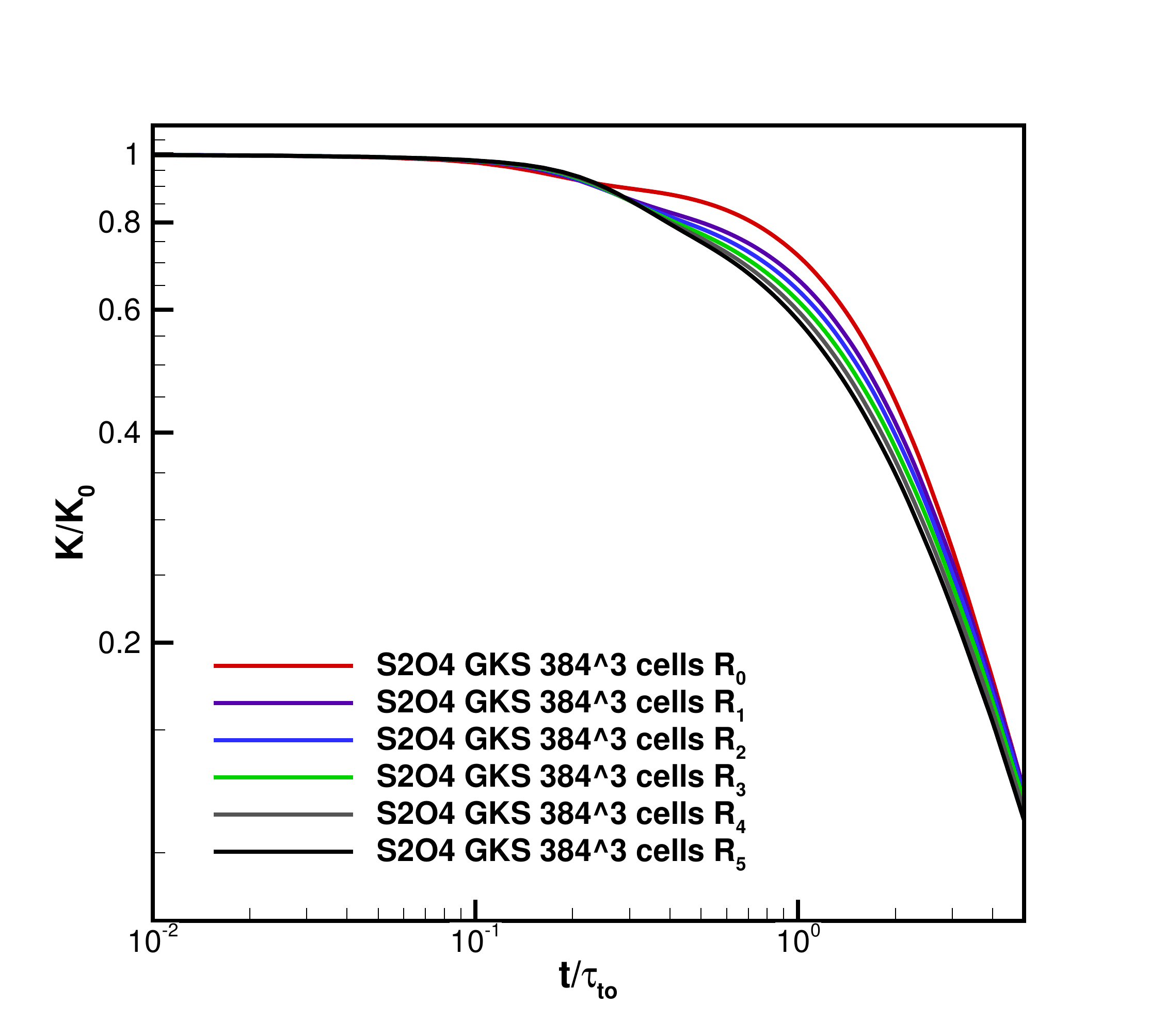}\\
\includegraphics[width=0.45\textwidth]{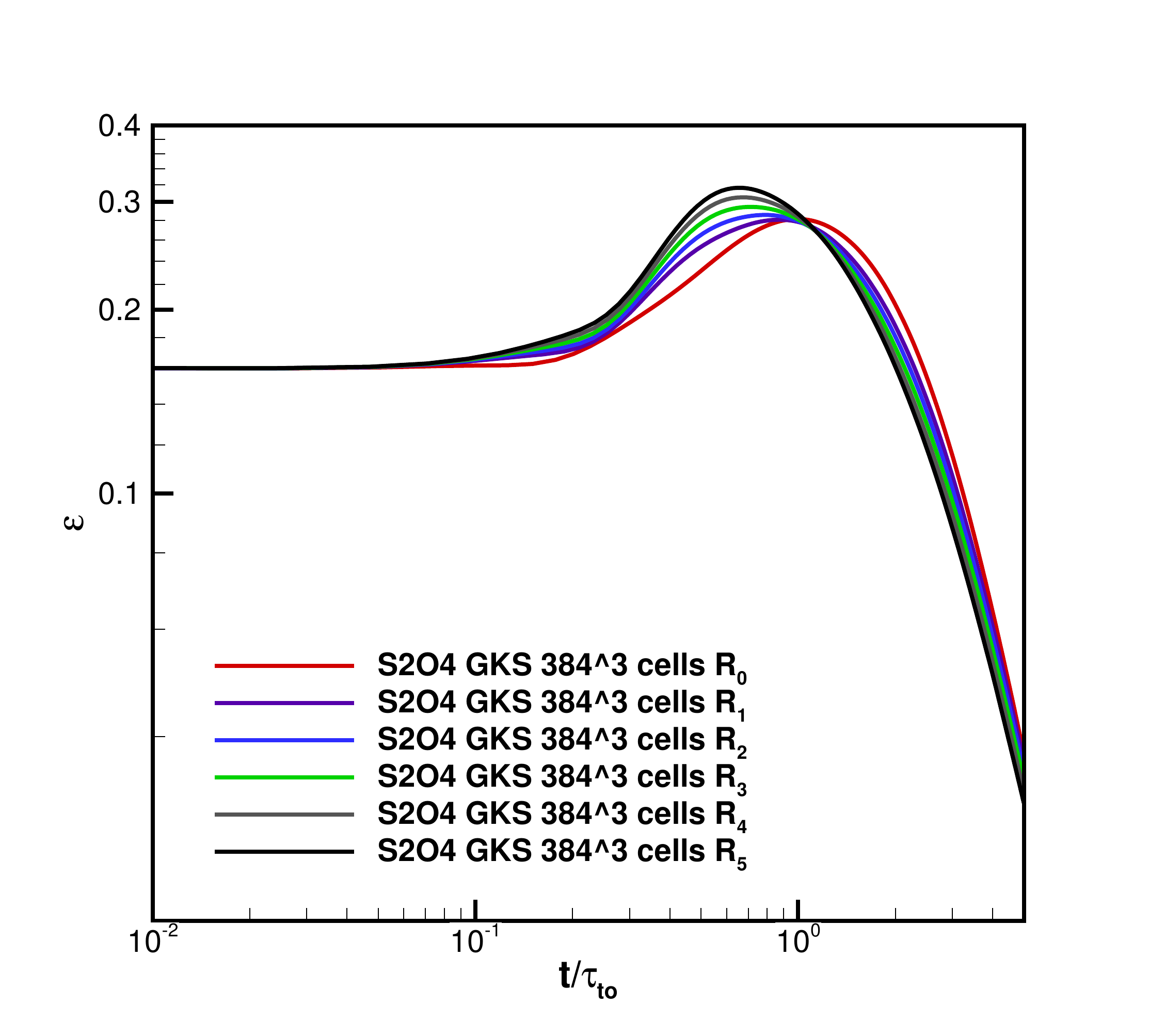}
\includegraphics[width=0.45\textwidth]{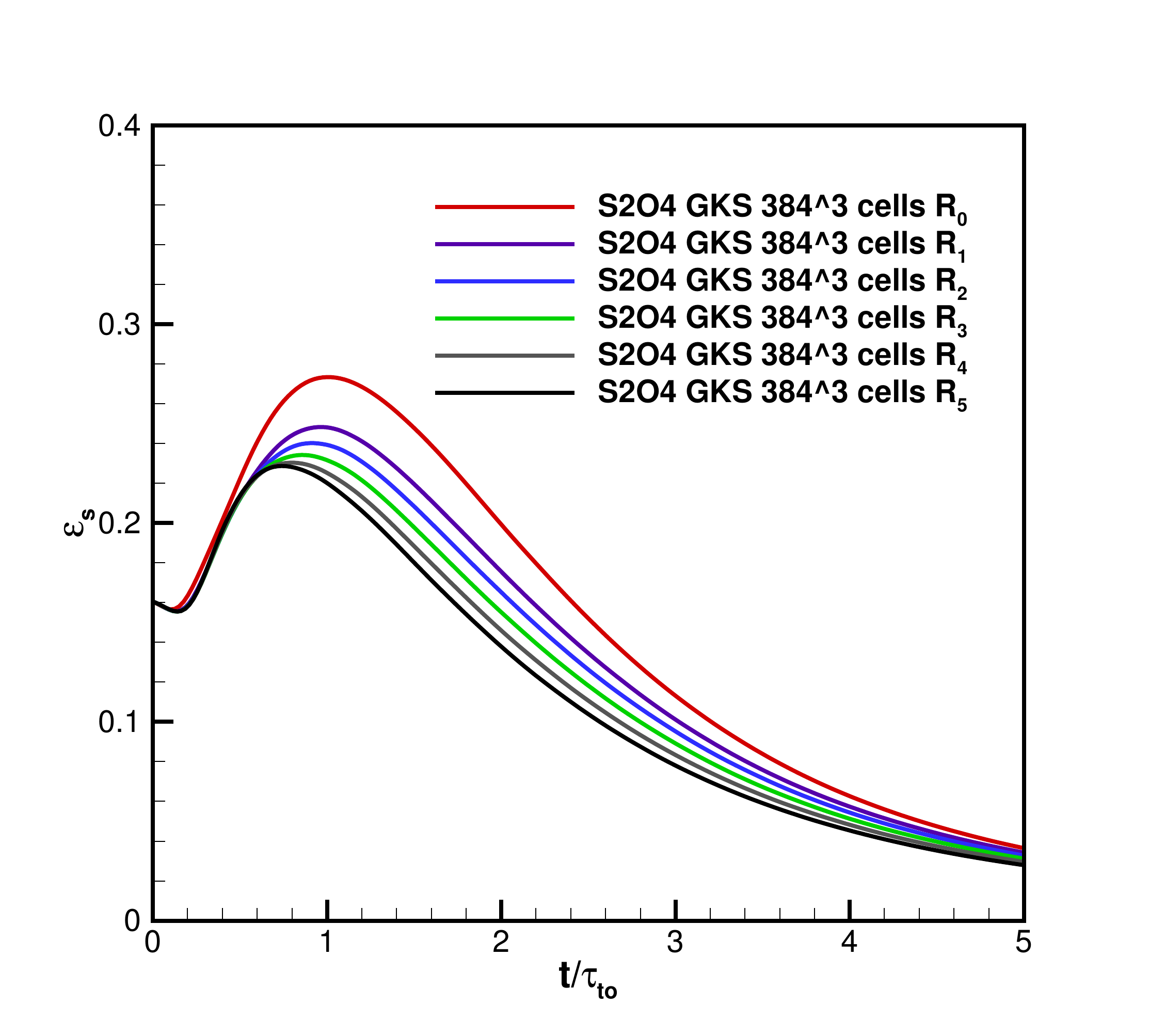}\\
\includegraphics[width=0.45\textwidth]{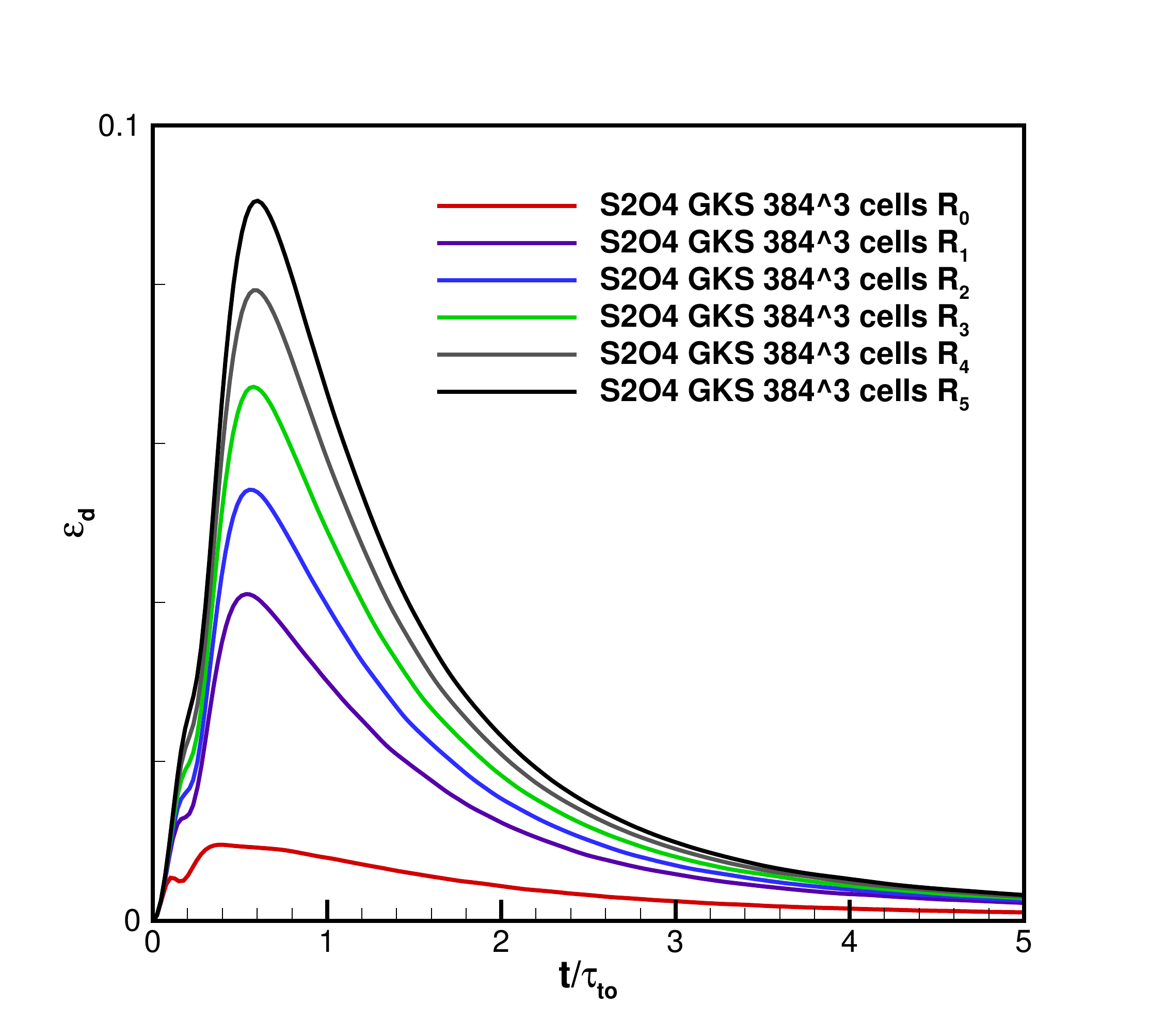}
\includegraphics[width=0.45\textwidth]{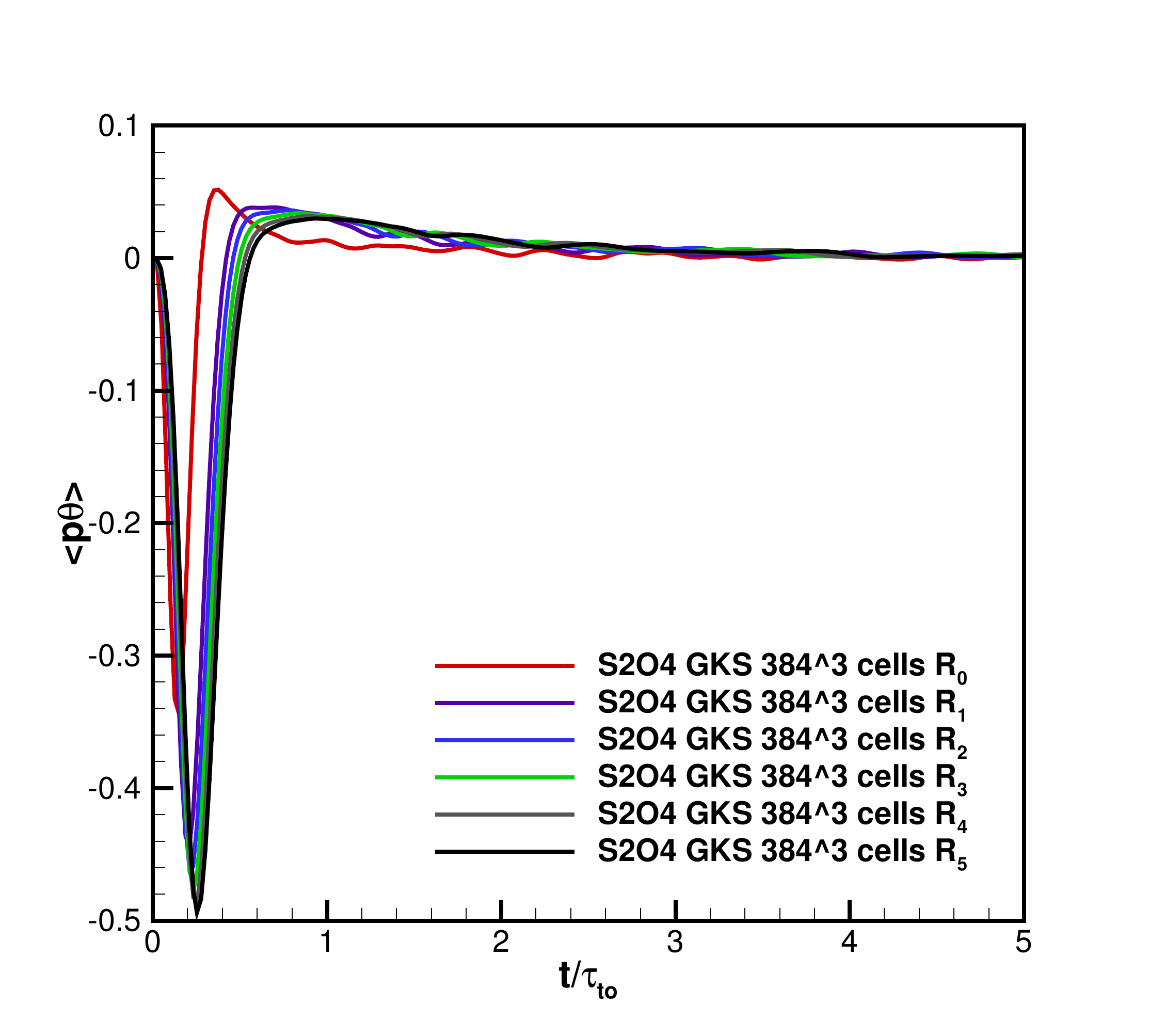}
\caption{\label{history_ma08_12} Time history of
$\rho_{rms}/Ma_{t}^2$, $K/K_0$, $\varepsilon$, $\varepsilon_d$,
$\varepsilon_s$ and $\left\langle p \theta \right\rangle$ for cases
$R_0$-$R_5$.}
\end{figure}

Time history of key statistical quantities are presented in
Fig.\ref{history_ma08_12}, which provides benchmark solution for
simulating isotropic compressible
turbulence up to supersonic regime. The normalized root-mean-square
density $\rho_{rms}/Ma_{t}^2$ decreases monotonically with the
increase of initial turbulent Mach number. As the initial turbulent
Mach number increases, the peak of dissipation increases as well.
For incompressible turbulence, the normalized turbulent kinetic
energy and ensemble total dissipation rate $\varepsilon$ are assumed
in universal power decaying rate as $K/K_0 \approx (t/t_0)^{-n}$ and
$\varepsilon/\varepsilon_0 \approx (t/t_0)^{-(n+1/n)}$, where $n$ is
usually treat as a constant, i.e. $n = 10/7$ based on the
Loitsianskii invariant \cite{sagaut2008homogeneous}, $n=6/5$
predicted assuming a constant Staffman invariant, and $n = 3/2$ in
\cite{saffman1967large}. However, as in subusonic regime \cite{samtaney2001direct}, it can be clearly observed that
the normalized turbulent kinetic energy and ensemble total
dissipation rate in isotropic compressible turbulence don't show any
universal power decaying rate. Obviously, ensemble solenoidal
dissipation rate $\varepsilon_s$ decreases with the increase of
$Ma_{t}$, while the dilational dissipation rate $\varepsilon_d$ 
rises with the increase of $Ma_{t}$. Remarkably, 
the peak of ensemble dilational dissipation rate $\varepsilon_d$ in
supersonic isotropic turbulence $Ma_{t} = 1.2$ is almost $8$ times
larger than that of subsonic isotropic turbulence $Ma_{t} = 0.5$.
The dilational dissipation mechanism has not been rarely absorbed in
traditional eddy-viscosity LES models \cite{xie2018modified}, and
current DNS results provide the first step results for 
constructing turbulence model in supersonic regime. The time of
lowest peak of ensemble pressure-dilation transfer $\left\langle p
\theta \right\rangle$ becomes larger with the higher initial
turbulence Mach number. In addition, $\left\langle p \theta
\right\rangle$ change signs during the evolution and preserve small
but positive thereafter, which agree with earlier study for subsonic
isotropic turbulence \cite{sarkar1991analysis}. It is reported that
the ratio between the ensemble pressure-dilation term and the right hand side
of Eq.(\ref{dkdt}) becomes small for solenoidal forced
quasi-stationary supersonic isotropic turbulence
\cite{wang2018kinetic}. However, during the early stage of the
decaying supersonic isotropic turbulence, the ensemble
pressure-dilation term can be in the same order of ensemble total
dissipation rate.

\begin{table}[!h]
\begin{center}
\def\temptablewidth{0.75\textwidth}
{\rule{\temptablewidth}{1.0pt}}
\begin{tabular*}{\temptablewidth}{@{\extracolsep{\fill}}c|cccc}
Test                          & $R_6$      &$R_7$         &$R_8$       &$R_9$  \\
\hline
$Re_{\lambda}$                &60           &40           &20          &10 \\
Grid size                     &$384^3$      &$384^3$      &$256^3$     &$256^3$  \\
$\kappa_{max}\eta_0$          &$2.97$       &$3.64$       &$3.43$      &$4.85$ \\
$\text{d}t_{ini}/\tau_{t_0}$  &$3.92/1000$ &$3.92/1000$
&$4.89/1000$ &$4.89/1000$
\end{tabular*}
{\rule{\temptablewidth}{1.0pt}}
\end{center}
\vspace{-5mm} \caption{\label{ma12gridtable} Minimum spatial
resolution $\kappa_{max} \eta_0$ for isotropic compressible
turbulence with different $Re_{\lambda}$. }
\end{table}

\begin{figure}[!h]
\centering
\includegraphics[width=0.49\textwidth]{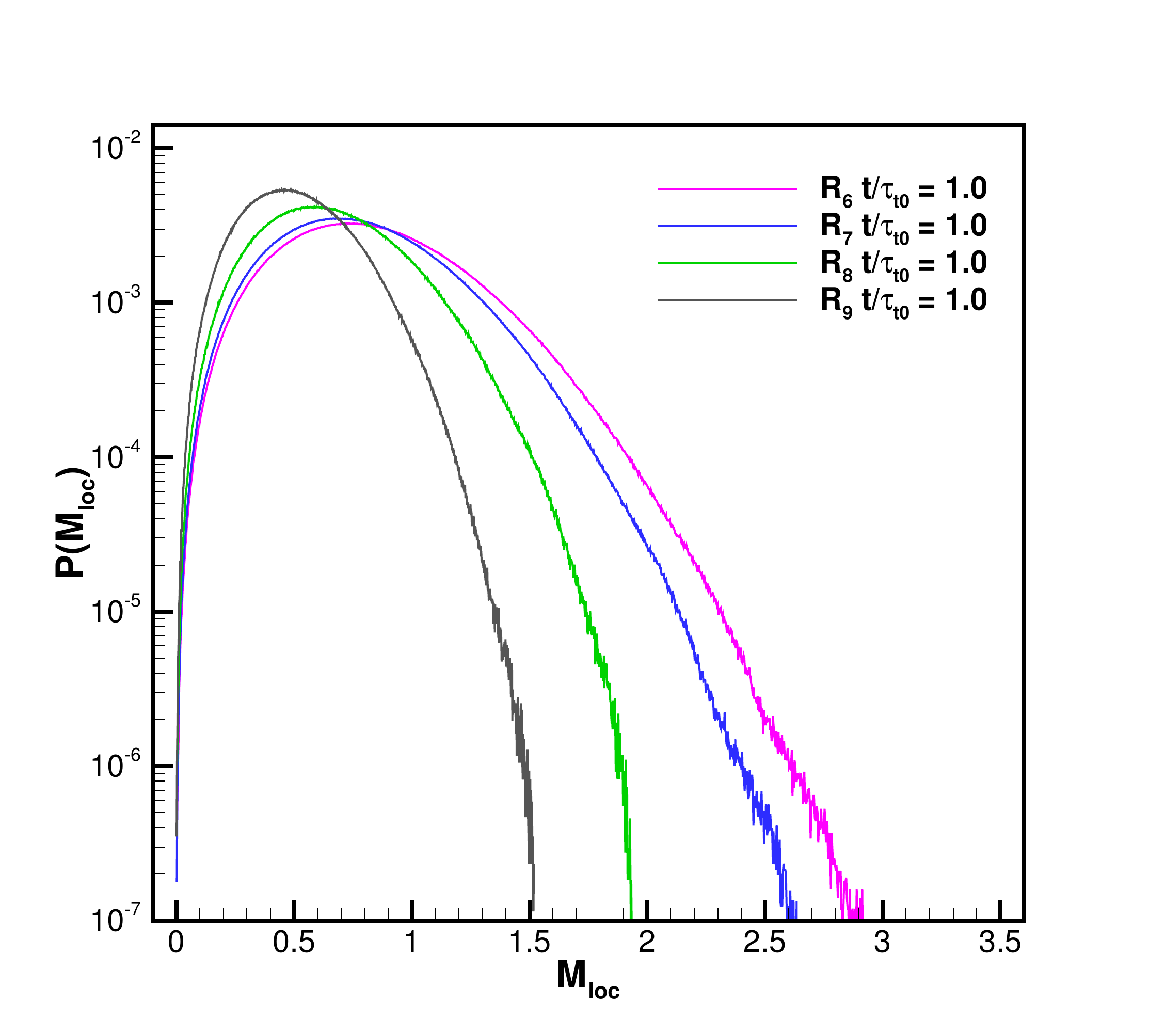}
\includegraphics[width=0.49\textwidth]{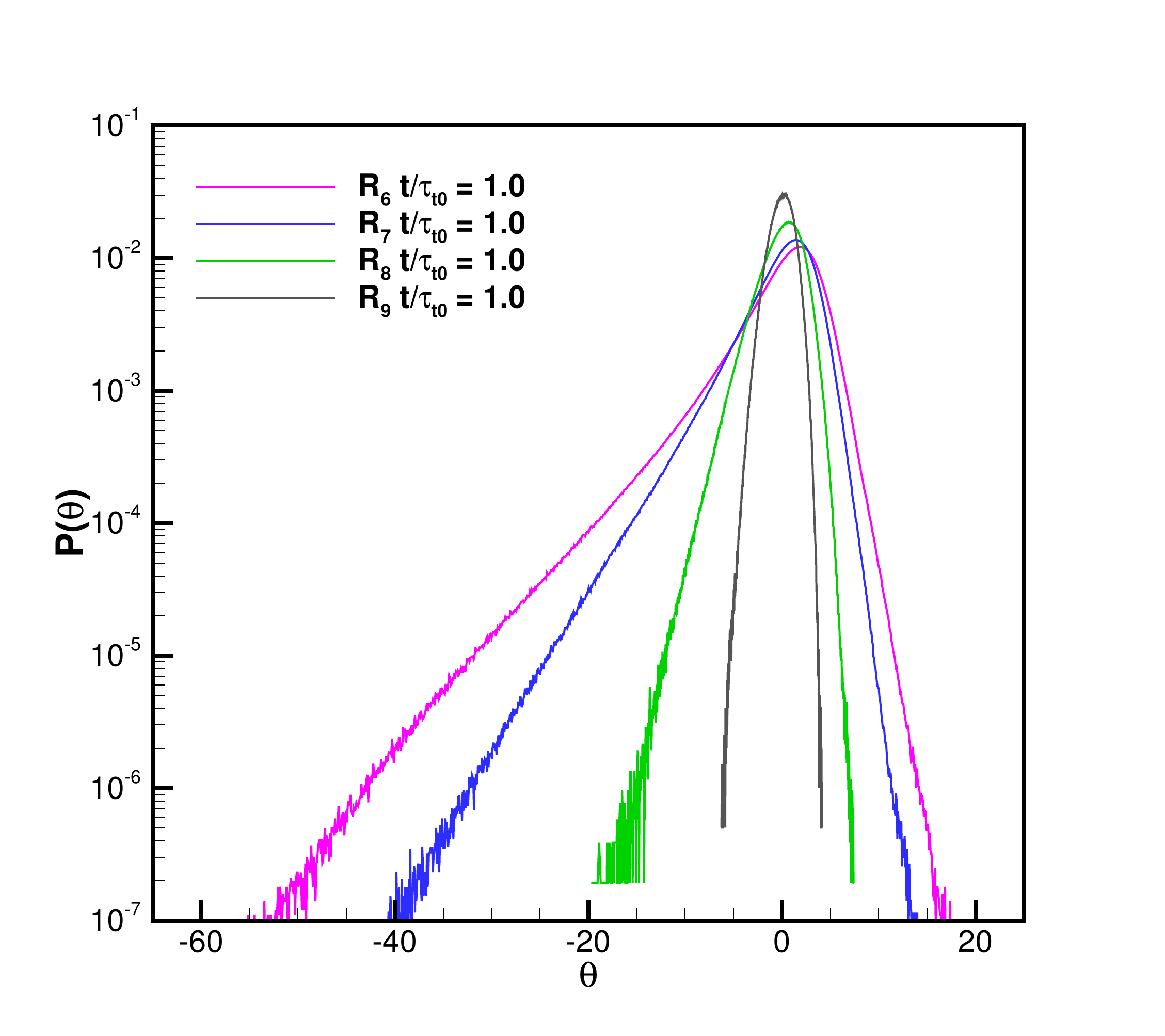}
\caption{\label{pdf_mloc_theta_re} PDF of local turbulent Mach
number and dilation at $t/\tau_{t_0} = 1.0$ of $R_6$, $R_7$, $R_8$
and $R_9$.} \centering
\includegraphics[width=0.49\textwidth]{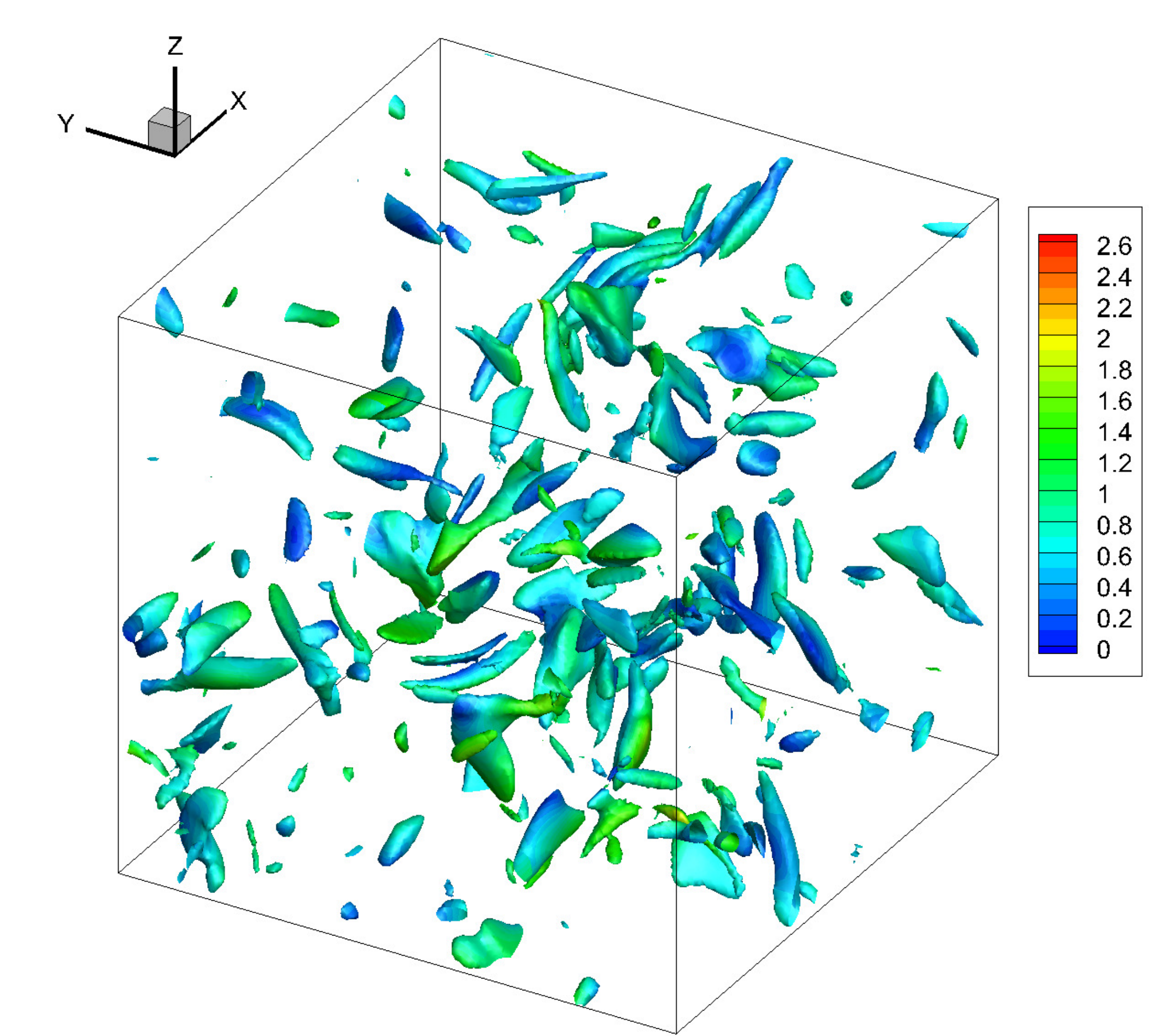}
\includegraphics[width=0.49\textwidth]{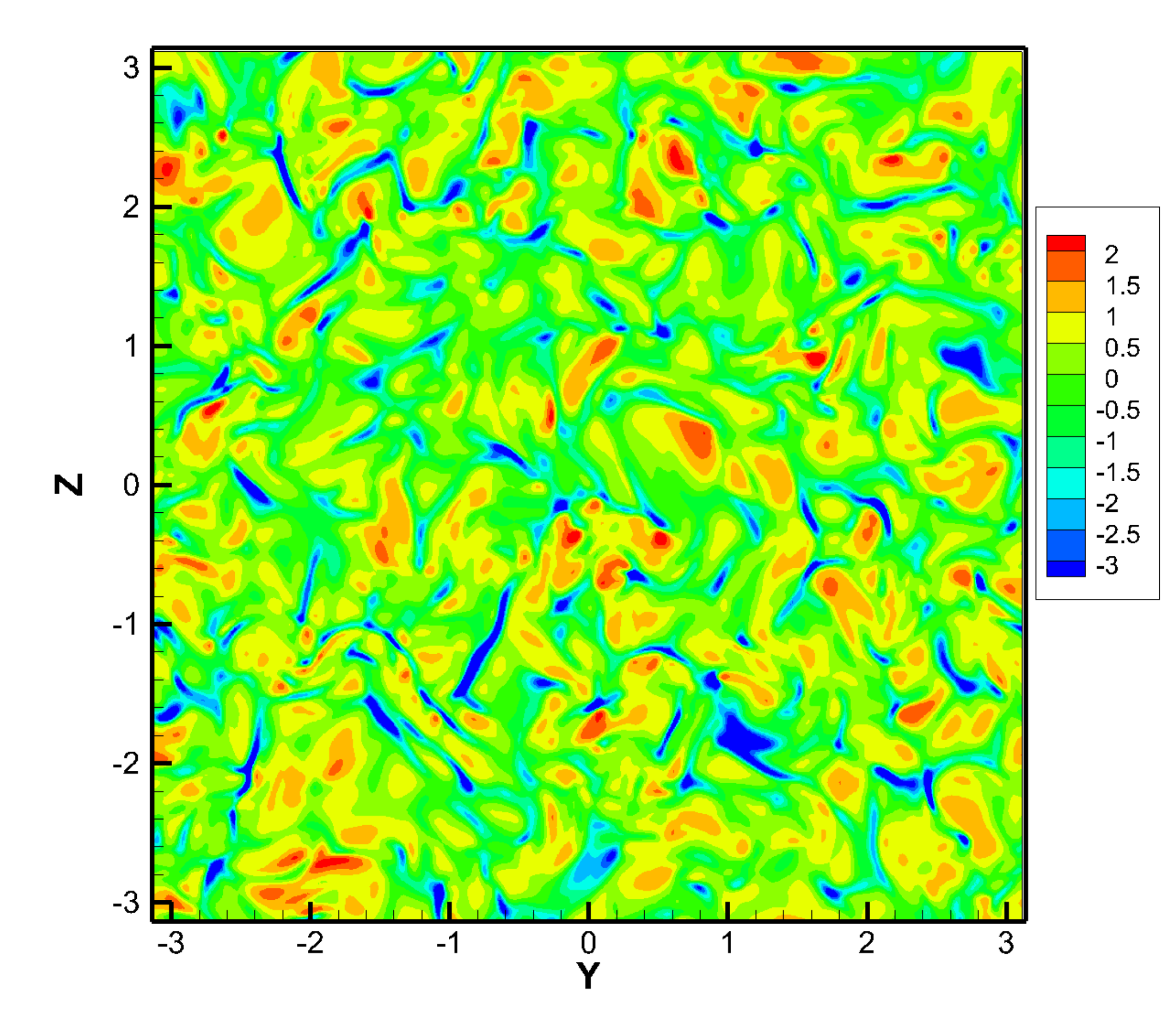}
\caption{\label{Re_effect_Q_theta40} Iso-surface of the second
invariant of velocity gradient tensor $Q = 25$ and contour of
normalized dilation $\theta/\left\langle \theta
\right\rangle^{\ast}$ for $R_7$ on $x = 0$ slice at $t/\tau_{t_0} =
1.0$.}
\end{figure}

\subsection{Taylor microscale Reynolds number effect}
In this section, the effect from the Taylor microscale Reynolds number for isotropic
compressible turbulence is studied. Power-law decay for
incompressible turbulence with low Taylor microscale Reynolds
number, i.e. $Re_{\lambda_0}\leq 50$ has been investigated in
earlier work \cite{huang1994power, mansour1994decay}. The current
study focuses on the isotropic turbulence with low Taylor microscale
Reynolds number in supersonic regime, and the cases $R_6-R_9$ with a
fixed supersonic turbulent Mach number $Ma_{t}=1.2$ are listed in
Table.\ref{ma12gridtable}. The grid size is set to meet the
requirement $\kappa_{max}\eta_0\ge 2.71$ and all simulations are
guided by previous criterion of fourth-order GKS. As shown in
Fig.\ref{pdf_mloc_theta_re}, PDFs of local turbulent Mach number
still show large portion of supersonic state at $t/\tau_{t_0} = 1.0$,
while the range of PDF decreases with the decreasing of Taylor
microscale Reynolds number. Meanwhile, PDFs of dilation presented in
Fig.\ref{pdf_mloc_theta_re} are skewed and the negative tails
resulted from the shocklets becomes shorter at the smaller Taylor
microscale Reynolds number. Isotropic compressible turbulence at 
lower Taylor microscale Reynolds number demonstrates a smaller range
of dilation, which is consistent with physical insight that the
lower Taylor microscale Reynolds number means the stronger viscous
effect, and the stronger dissipation smooths the flow fields and leads
to weaker compression regions and expansion regions. Iso-surface of the
second invariant of velocity gradient tensor $Q = 25$ and contour of
normalized dilation $\theta/\left\langle \theta
\right\rangle^{\ast}$ for $R_7$ at $x= 0$ slice are presented in
Fig.\ref{Re_effect_Q_theta40}, where $\left\langle \theta
\right\rangle^{\ast}$ is root-mean-square dilation at
$t/\tau_{t_0}=1.0$. Iso-surface is colored by local turbulent Mach
number in a $128^3$ sub-domain covering $(78 \eta_0)^3$, where the
sub-domain is located at the center of the full domain. A smaller
range of vortex structure in flow fields is observed in
Fig.\ref{Re_effect_Q_theta40} compared with
Fig.\ref{Ma12_effect_Q_theta}. Contour of normalized dilation
$\theta/\left\langle \theta \right\rangle^{\ast}$ shows the wider
`ribbon'  for strong shocklets and bigger `block' for high expansion
region compared with these from moderate Taylor microscale Reynolds
number. This result confirms the fact that the stronger dissipation for 
the lower Taylor microscale Reynolds number supersonic compressible
turbulence smooths the flow fields and makes the
transition of compression regions gently.

\begin{figure}[!h]
\centering
\includegraphics[width=0.45\textwidth]{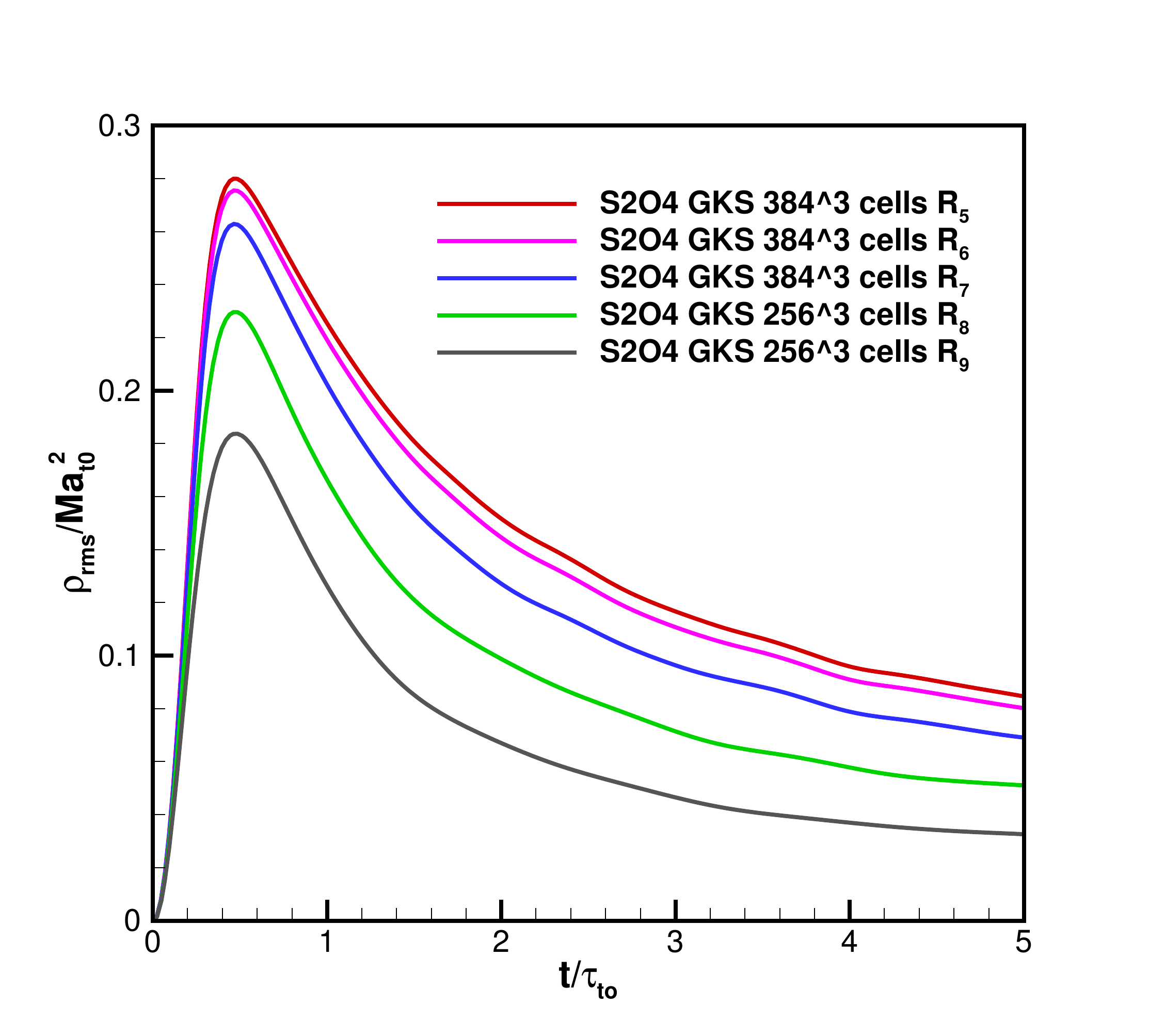}
\includegraphics[width=0.45\textwidth]{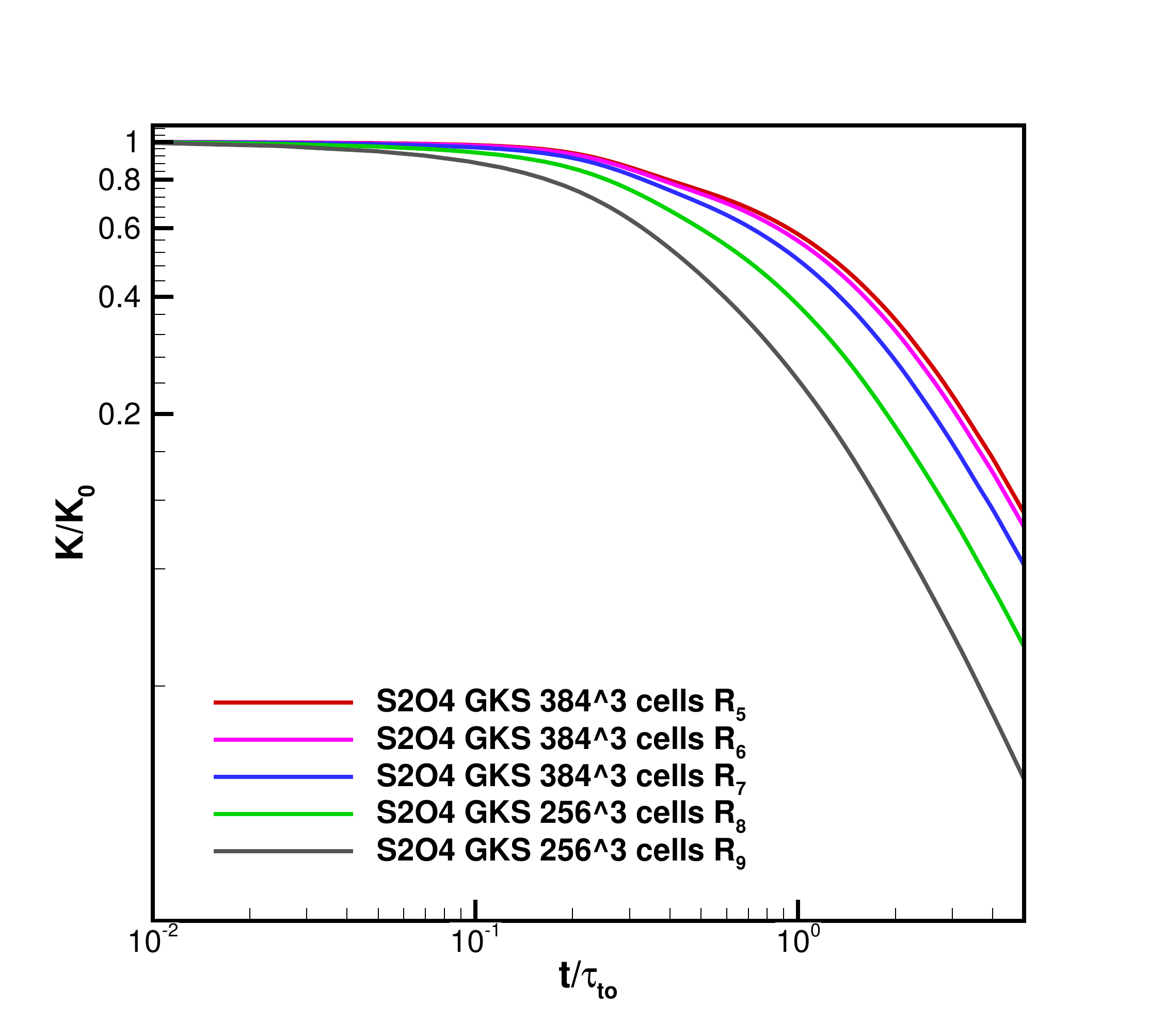}\\
\includegraphics[width=0.45\textwidth]{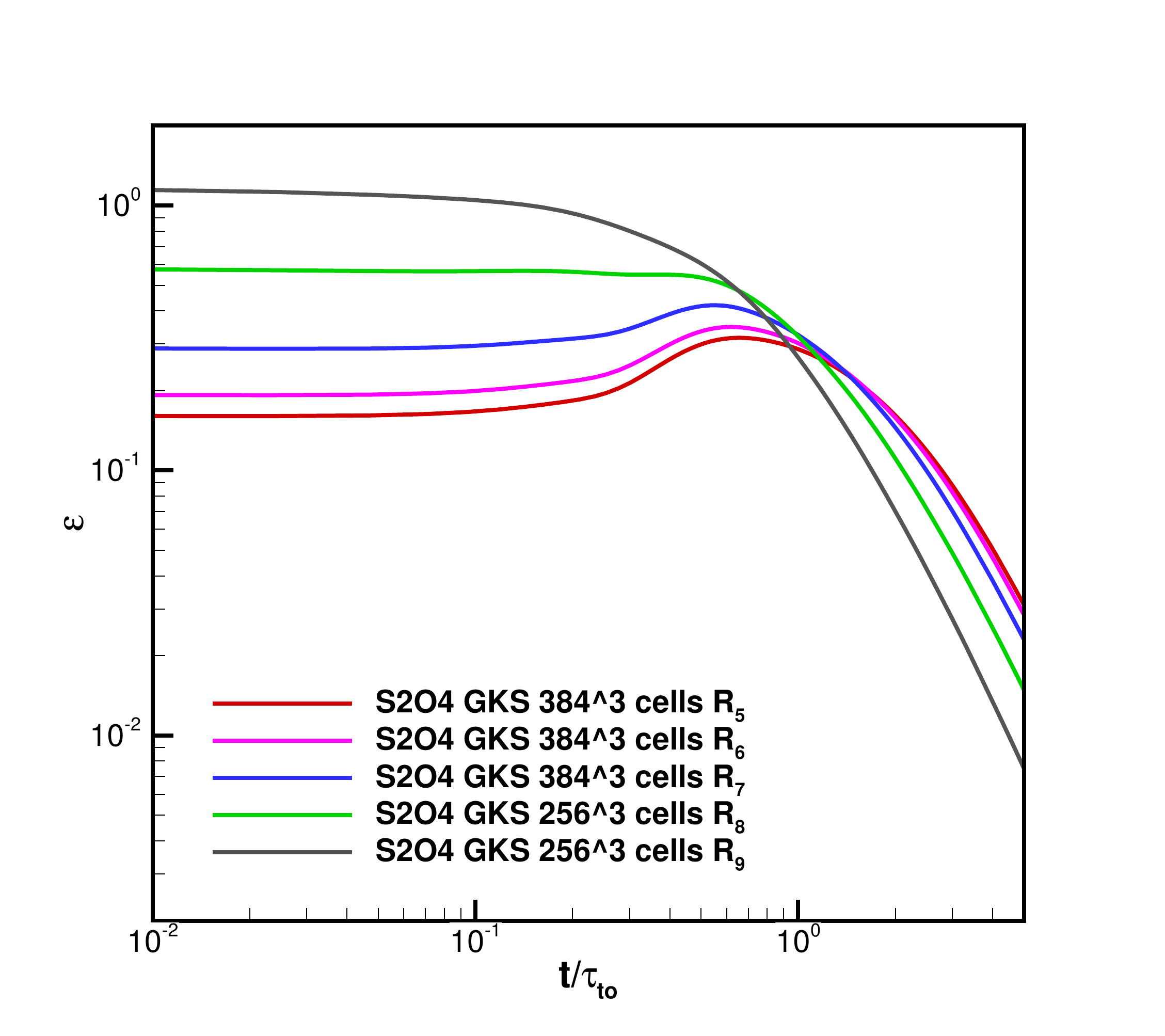}
\includegraphics[width=0.45\textwidth]{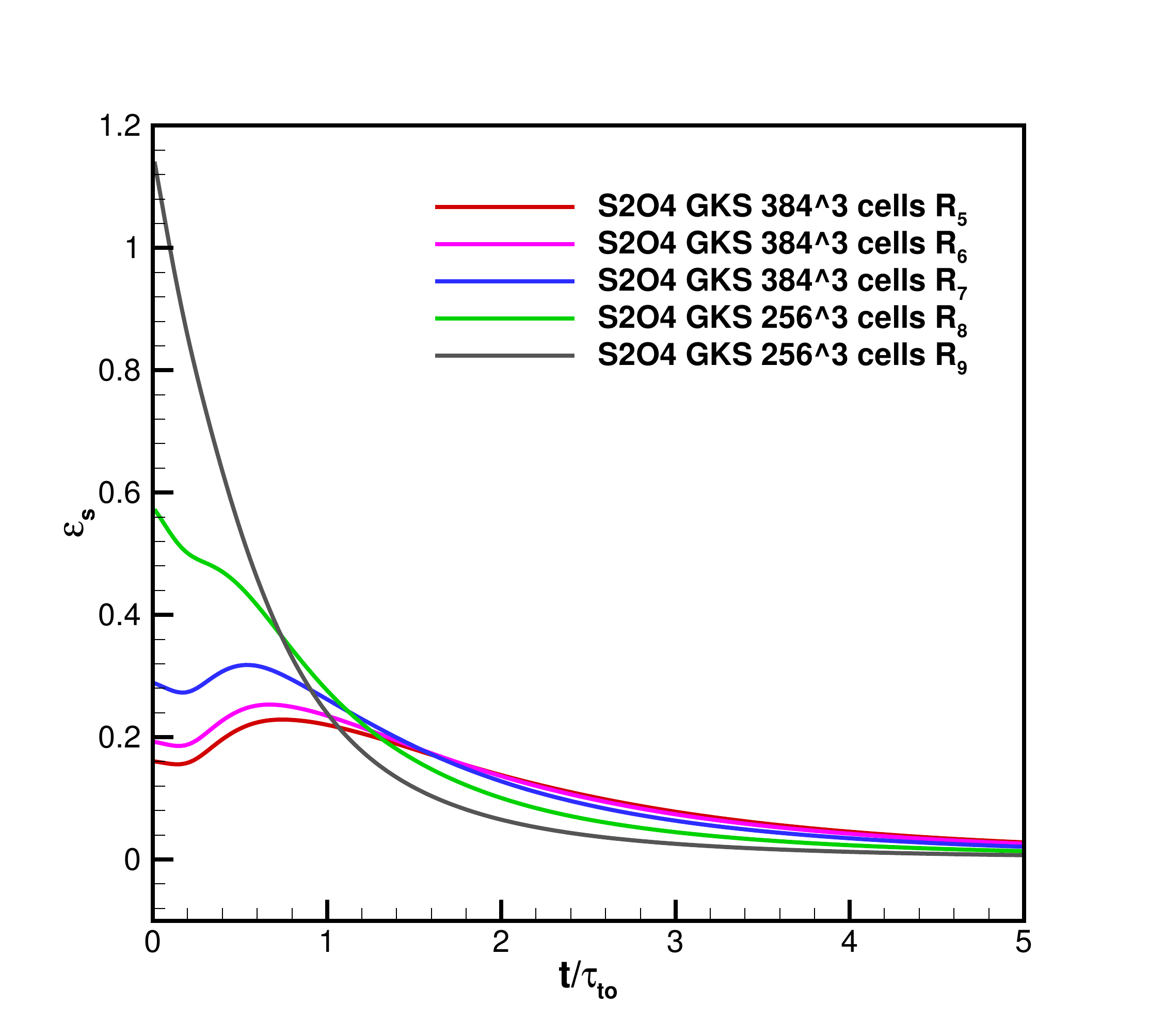}\\
\includegraphics[width=0.45\textwidth]{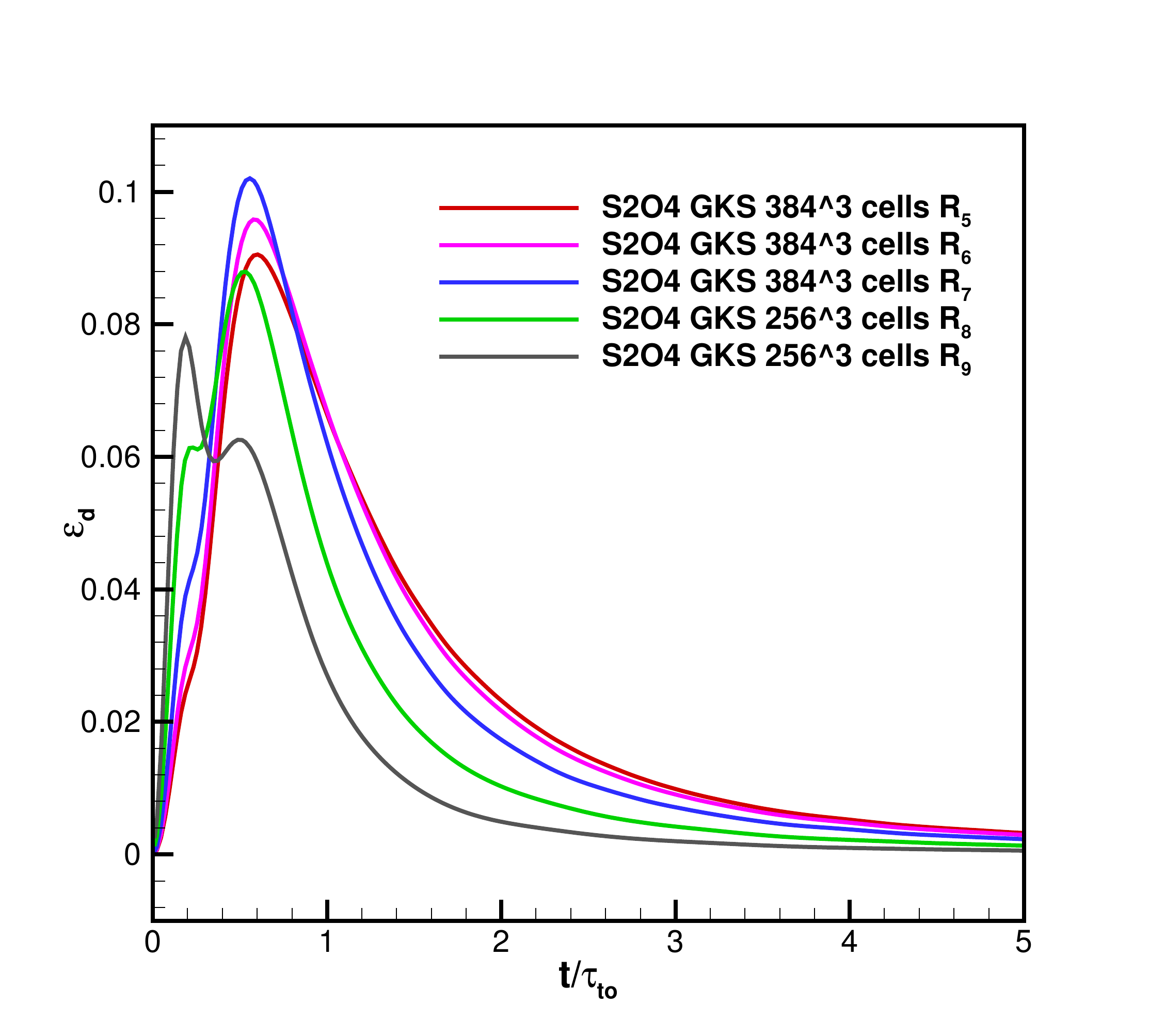}
\includegraphics[width=0.45\textwidth]{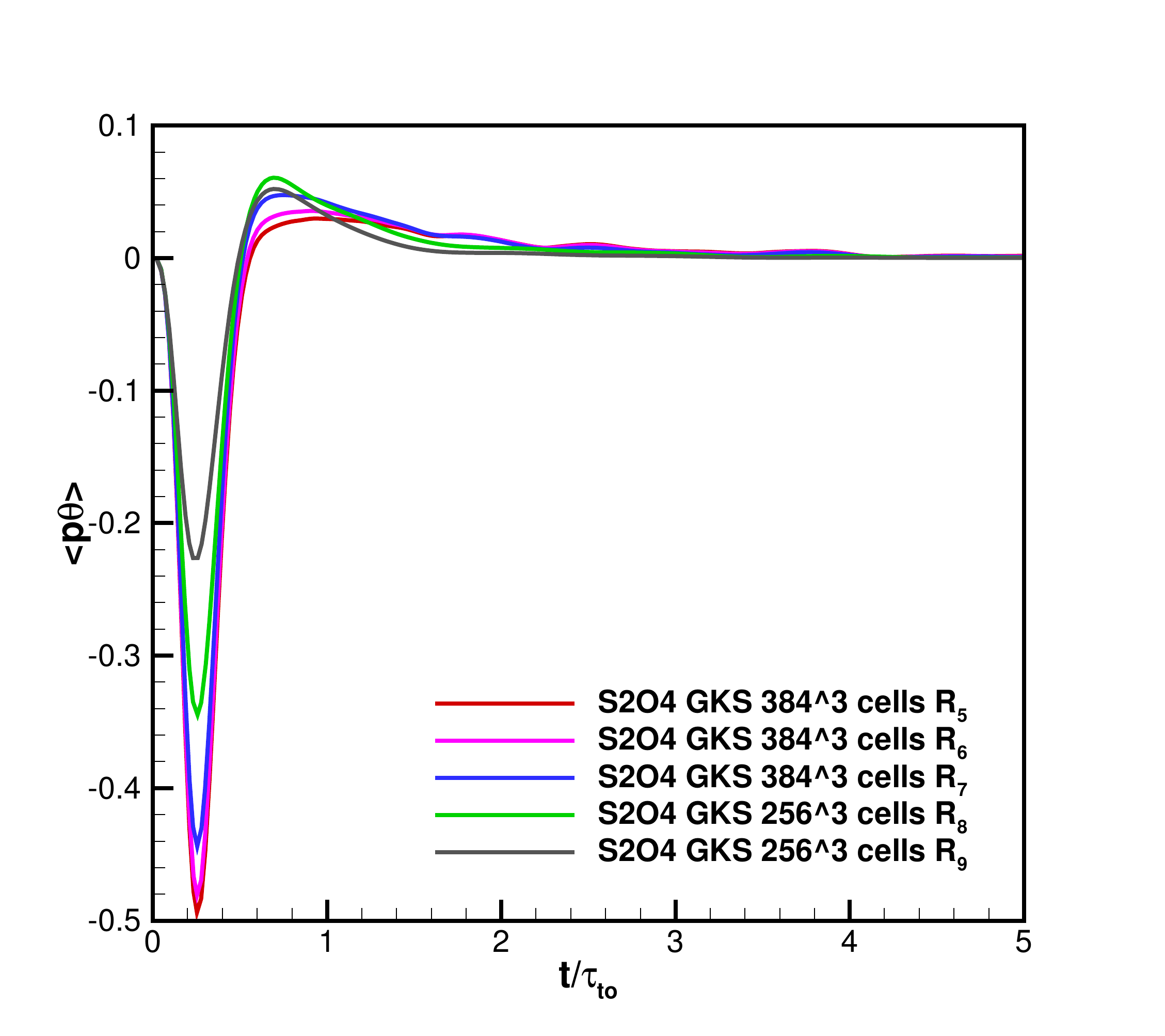}
\caption{\label{history_re20_60} Time history of
$\rho_{rms}/Ma_{t0}^2$, $K/K_0$, $\varepsilon$, $\varepsilon_d$,
$\varepsilon_s$ and $\left\langle p \theta \right\rangle$ for cases
$R_6$-$R_9$.}
\end{figure}

The key statistical quantities are presented in
Fig.\ref{history_re20_60} which provide benchmark solutions for studying
supersonic isotropic turbulence at low Taylor microscale Reynolds
number. The normalized root-mean-square density
$\rho_{rms}/Ma_{t}^2$ decreases with the decrease of initial Taylor
microscale Reynold number. The lower Taylor microscale Reynolds
number corresponds to  higher ensemble total dissipation rate
$\varepsilon$ during the evolution of this system, which is the
direct result from the stronger viscous effect at lower
$Re_{\lambda}$. It is clear that ensemble solenoidal dissipation
rate $\varepsilon_s$ decreases with the decrease of
$Re_{\lambda}$, while the ensemble dilational dissipation rate
$\varepsilon_d$ seems slightly dependent on $Re_{\lambda}$. The
behavior of ensemble dilational dissipation rate needs further study
in detail, which is pretty meaningful for constructing compressible
LES model for supersonic and hypersonic compressible turbulence. In addition, the solenoidal and dilational dissipation rate play the dominant role in hypersonic transition to turbulence reported in previous experiments \cite{lee2008tran,zhu2016tran,zhu2018aero}. Current high-order robust scheme will be used to validate and provide more detailed analysis for such hypersonic flows. The
moment of lowest peak of ensemble pressure-dilation transfer
$\left\langle p \theta \right\rangle$ is independent of
$Re_{\lambda}$ and only depends on $Ma_{t}$ compared with
Fig.\ref{history_ma08_12}, while the lowest peak decreases with the
decrease of $Re_{\lambda}$. Similar with Fig.\ref{history_ma08_12},
$\left\langle p \theta \right\rangle$ changes signs during the
evolution and preserves small but positive value. The similar
behavior is observed that the ensemble pressure-dilation term can
not be neglected compared with ensemble total dissipation rate in
the early stage of evolution, as it has the same order of ensemble
total dissipation rate.

\section{Conclusions}

This paper intends to address the accuracy and robustness of HGKS in
DNS for isotropic compressible turbulence simulations up to supersonic regime.
Key statistical quantities are compared with high-order compact
finite difference scheme to determine the DNS criterion. As a balance between the 
robustness and accuracy, the WENO-Z
reconstruction is properly chosen in the current scheme. 
The numerical tests show that the
minimum spatial resolution parameter $\kappa_{max} \eta_0 \ge 2.71$
and the maximum temporal resolution parameter $\Delta
t_{ini}/\tau_{t_0} \leq 5.58/1000$ for the fourth-order GKS is adequate
for resolving the isotropic compressible turbulence. Guided by such
a criterion, isotropic compressible turbulence are simulated for turbulent Mach
number $Ma_t$ from the nonlinear subsonic regime $0.8$ to the
supersonic one $1.2$, and low Taylor microscale Reynolds number
from $10$ to $72$. 
A wide range for PDF of
local turbulent Mach number, strong random shocklets, and high
expansion regions appear with  high initial turbulent Mach
number. The isotropic turbulence with high turbulent Mach number
up to supersonic regime has been studied. The accuracy and robustness of the fourth-order GKS
have been fully confirmed. Statistical quantities are
provided for these cases, which provide benchmark solutions
for supersonic isotropic compressible turbulence. The ensemble
budget of the turbulent kinetic energy is analyzed, 
which plays an important data base in
modeling supersonic and hypersonic compressible turbulence. The
solenoidal dissipation rate decreases with the increase of both
$Ma_t$ and $Re_{\lambda}$. Meanwhile, the dilational dissipation
rate increases with the increase of $Ma_t$ due to strong compressibility effect, 
and it seems slightly dependent on $Re_{\lambda}$. 
The HGKS provides a valid tool for studying
compressible turbulence. The physics of isotropic compressible
turbulence as well as the construction of compressible LES model in
supersonic regime will be studied. At the current stage, the DNS on a much higher turbulent Mach
number up to $Ma_{t} = 2.2$ and higher Taylor microscale Reynolds
number $Re_{\lambda}=100$ have been obtained by HGKS. All these results and the analysis of physical mechanism of 
isotropic compressible turbulence will be presented in the subsequent paper.

\section*{Ackonwledgement}
The current research is supported by National Science Foundation of
China (11701038, 11772281, 91852114) and the Fundamental Research
Funds for the Central Universities. The authors would like to thank
TianHe-II in Guangzhou for providing high performance computational
resources.

\end{document}